\documentclass[twocolumn,showpacs,preprintnumbers,amsmath,amssymb]{revtex4}

\usepackage{graphicx}
\usepackage{dcolumn}
\usepackage{bm}

\begin{document}


\title{
Broadband method for precise microwave spectroscopy of superconducting thin films near the critical temperature
}

\author{Haruhisa Kitano}
 \altaffiliation[corresponding author: ]{hkitano@phys.aoyama.ac.jp}
 \affiliation{
 Department of Physics and Mathematics, Aoyama Gakuin University, 5-10-1 Fuchinobe, Sagamihara, Kanagawa 229-8558, Japan
 }
\author{Takeyoshi Ohashi}%
\author{Atsutaka Maeda}%
\affiliation{%
Department of Basic Science, University of Tokyo, 3-8-1, Komaba, Meguro-ku, Tokyo 153-8902, Japan
}%

\date{\today}

\begin{abstract}
We present a high-resolution microwave spectrometer to measure the frequency-dependent complex conductivity of a superconducting thin film near the critical temperature. 
The instrument is based on a broadband measurement of the complex reflection coefficient, $S_{\rm 11}$, of a coaxial transmission line, which is terminated to a thin film sample with the electrodes in a Corbino disk shape. 
In the vicinity of the critical temperature, the standard calibration technique using three known standards fails to extract the strong frequency dependence of the complex conductivity induced by the superconducting fluctuations. 
This is because a small unexpected difference between the phase parts of $S_{\rm 11}$ for a short and load standards gives rise to a large error in the detailed frequency dependence of the complex conductivity near the superconducting transition. 
We demonstrate that a new calibration procedure using the normal-state conductivity of a sample as a load standard resolves this difficulty. 
The high quality performance of this spectrometer, which covers the frequency range between 0.1~GHz and 10~GHz, the temperature range down to 10~K, and the magnetic field range up to 1~T, is illustrated by the experimental results on several thin films of both conventional and high temperature superconductors. 

\end{abstract}

\maketitle


\section{\label{sec:intro}Introduction}
The cavity perturbation method using a microwave resonator has been played a central role in the investigation of the electrodynamic properties of unknown materials in the microwave and millimeter wave regions \cite{GrunerReview1,GrunerReview2,GrunerReview3}. 
Particularly, the development of a superconducting resonator with a high-quality-factor (high-$Q$) enabled a high-resolution measurement of the complex microwave conductivity of a tiny single crystal in zero magnetic field \cite{Sridhar1988,Bonn1991}. 
Very recently, a rutile dielectric resonator with $Q$ factors in excess of 10$^6$ has been developed, which can be applied to a similar high-resolution measurement under magnetic field \cite{Huttema2006}. 
In addition, the cavity perturbation method becomes a powerful tool to probe the electric conduction of unknown powder samples, where both the dc resistivity measurement and the optical measurement are difficult to be performed \cite{KitanoPRL2002,KitanoJPSJ2006}. 

However, it is almost impossible to obtain the microwave conductivity as a continuous function of frequency by using the cavity perturbation method. 
This is because a high stability and a high reproducibility of the resonant frequency, which is indispensable to the cavity perturbation technique, is no longer achieved by a continuous sweep of frequency. 
Although the information on the complex conductivity at several frequencies can be obtained by the use of several resonators with different resonant frequencies \cite{Hosseini1999,KitanoEPL2001}, the broadband microwave spectroscopy requiring the full information on the frequency dependence cannot be realized by this method. 
Thus, a complementary broadband method is required. The pioneering work of this technique was done by Booth {\it et al.} \cite{Booth1994}. 
They succeeded in measuring the complex conductivity of a thin film of high temperature superconductor as a continuous function of frequency between 45~MHz and 45~GHz at liquid nitrogen temperatures \cite{Booth1996}. 
The use of a nichrome thin film resistor as a load standard enabled a better calibration of long lossy transmission lines, which were often used in the cryogenic systems down to 0.1~K \cite{Stutzman2000}. 

In this article, we describe an apparatus designed for the microwave spectrometer to measure the complex conductivity of a highly conductive thin film sample as a function of frequency \cite{KitanoPhysicaC}. 
In particular, we focus on the precise measurement of the frequency-dependent complex conductivity of  a superconducting thin film near the critical temperature, $T_c$ \cite{KitanoPRB2006,OhashiPRB2006}. 
In the vicinity of $T_c$, we found that the standard calibration technique using three known standards was insufficient to obtain the frequency-dependent excess parts of the complex conductivity, which were induced by the superconducting fluctuations. 
We demonstrate that a new calibration procedure using the normal-state conductivity of a sample as a load standard resolves this difficulty. 
This spectrometer covers the frequency range between 0.1~GHz and 10~GHz, the temperature range from room temperature to 10~K, and the magnetic field range up to 1~T. 
The performance of it is illustrated by the experimental results on several thin films of both conventional and high temperature superconductors. 


\section{\label{sec:broadband}Broadband microwave measurement}
In a one-port reflection geometry, the load impedance $Z_L$ of a sample under test, which is terminated to the end of a transmission line, is related to the complex reflection coefficient $S_{\rm 11}$, as follows \cite{Pozar}: 
\begin{equation}
Z_L=Z_c\frac{1+S_{\rm 11}}{1-S_{\rm 11}},
\label{eq:ZL}
\end{equation}
where $Z_c$ is the characteristic impedance of the transmission line (usually, 50~$\Omega$). 
In the case of a coaxial line, $Z_c$ is given by $\Gamma Z_0/\sqrt{\epsilon_r}$, where $Z_0(=377~\Omega$) is the impedance of free space, and $\epsilon_r$ is the relative permittivity of a dielectric part in the coaxial line. 
$\Gamma$ is a geometrical factor given by $\ln(b/a)/2\pi$, where $a$ and $b$ are the inner and outer radius of the coaxial system, respectively. 
When the sample is a highly conductive thin film, $Z_L$ is given by 
\begin{equation}
Z_L=\Gamma Z_S\coth (kt),
\label{eq:Zs_eff}
\end{equation}
where $Z_S$, $k$, $t$ are the surface impedance of a bulk material, the complex propagation constant in the material, and the film thickness, respectively \cite{Booth1994}. 
If the film thickness $t$ is sufficiently smaller than the magnitude of the complex skin depth, which is given by $1/|k|$, we can approximate $\coth(kt)$ by $1/kt$. 
Thus, we obtain the following simple relationship between $Z_L$ and the complex conductivity, $\sigma(\equiv\sigma_1-i\sigma_2)$, of the material: 
\begin{equation}
Z_L\approx\frac{\Gamma}{\sigma t} ,
\label{eq:ZL_film}
\end{equation}
by using the expression of $Z_S(\equiv\sqrt{i\mu_0\omega/\sigma})$ in the local electrodynamics and that of $k(\equiv\sqrt{i\mu_0\omega\sigma})$, 
where $\mu_0$ and $\omega$ are the permiability of free space and the angular frequency, respectively. 

As will be described in the next section, we used a commercial 2.4~mm coaxial adapter \cite{OS2.4}, where the dielectric part is air (or vacuum), in order to terminate the thin film sample. 
In this case, $\epsilon_r$ can be regarded as almost unity. 
Therefore, we can obtain the complex conductivity of a highly conductive thin film sample as a function of frequency from the one-port reflection measurement of $S_{\rm 11}(\omega)$, as follows: 
\begin{equation}
\sigma(\omega)=\frac{1}{tZ_0}\frac{1-S_{11}(\omega)}{1+S_{11}(\omega)}.
\label{eq:sigma}
\end{equation}

It is important to note that Eqs.~(\ref{eq:Zs_eff})$\sim$(\ref{eq:sigma}) are derived under the assumption that TEM waves are dominantly propagated into the sample. 
This assumption is valid as long as the system can be regarded as a load-ended coaxial line. 
On the other hand, in the case of a open-ended coaxial line \cite{Jiang1993,Martens2000}, we must consider the influence of TM modes excited in a dielectric sample more precisely. 
This means that Eqs.~(\ref{eq:Zs_eff})$\sim$(\ref{eq:sigma}) cannot be applied when the influence of a dielectric substrate is not negligible. 
The contribution from the substrate is discussed 
in more detail in Appendix A.

\section{\label{sec:apparatus}Measurement apparatus}

\begin{figure}[b]
\includegraphics[height=5.1cm,width=8.5cm]{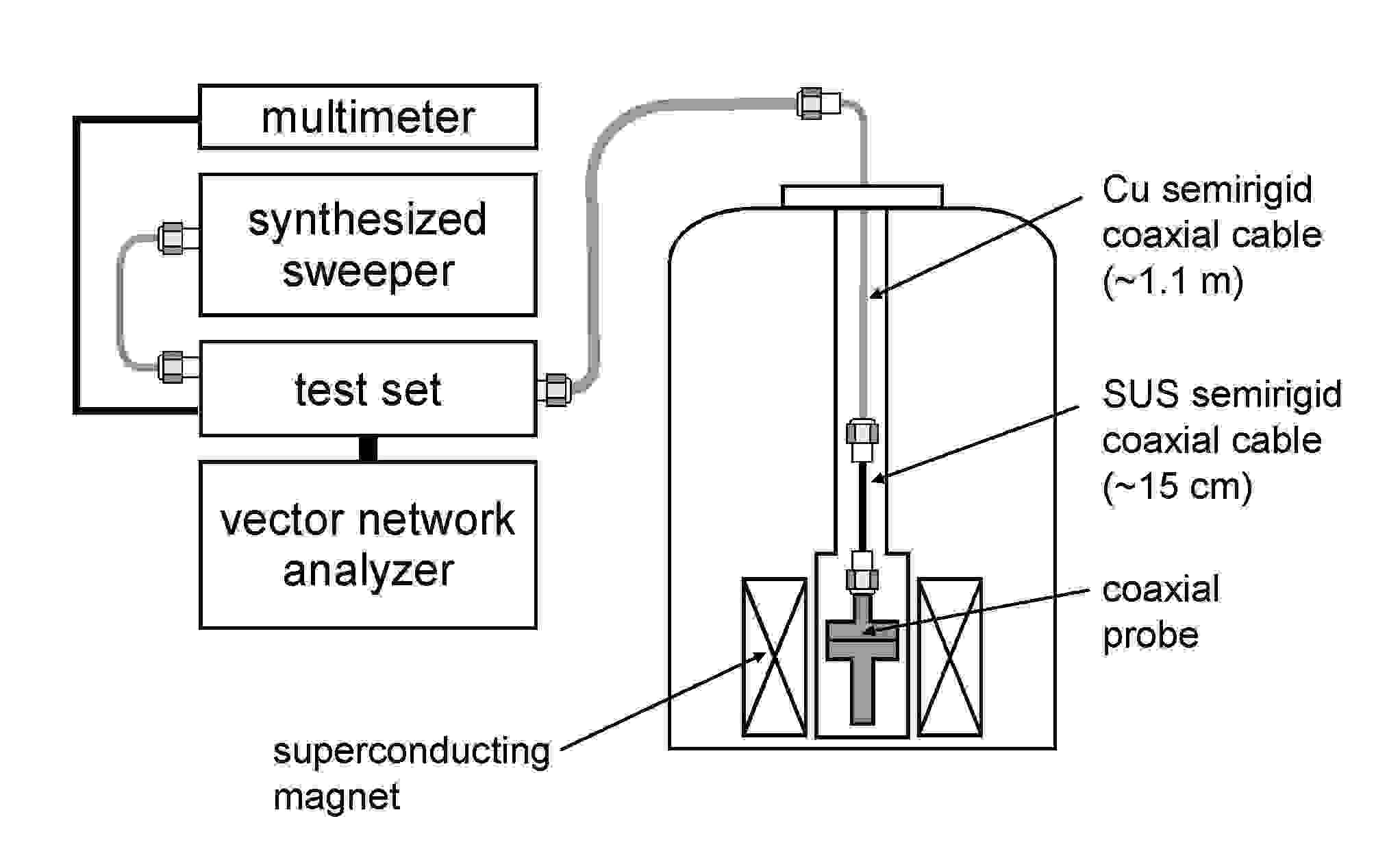}
\caption{
Scheme of microwave broadband measurement system. 
}
\label{fig:scheme}
\end{figure}

Figure \ref{fig:scheme} shows a schematic drawing of our experimental configuration. 
The microwave signal powered by a Hewlett-Packard (HP) 83650B synthesized sweeper is delivered to a thin film sample through an HP8517B $S$-parameter test set and a low temperature apparatus. 
The low temperature apparatus involves a coaxial assembly made of approximately 1.1~m of 0.085 in. copper coaxial cable and 15~cm of 0.085 in. stainless-steel coaxial cable \cite{coax_cable}. 
The reflected signal is measured over the frequency range from 45~MHz to 50~GHz using a HP8510C vector network analyzer. 
The total length of the coaxial lines from the network analyzer to the sample is about 1.7~m, and the whole loss of coaxial lines is about 5~dB at 10~GHz. 
The network analyzer is operated in step-sweep mode to ensure the phase coherence at each frequency. 

\begin{figure}[t]
\includegraphics[height=11.5cm,width=8.0cm]{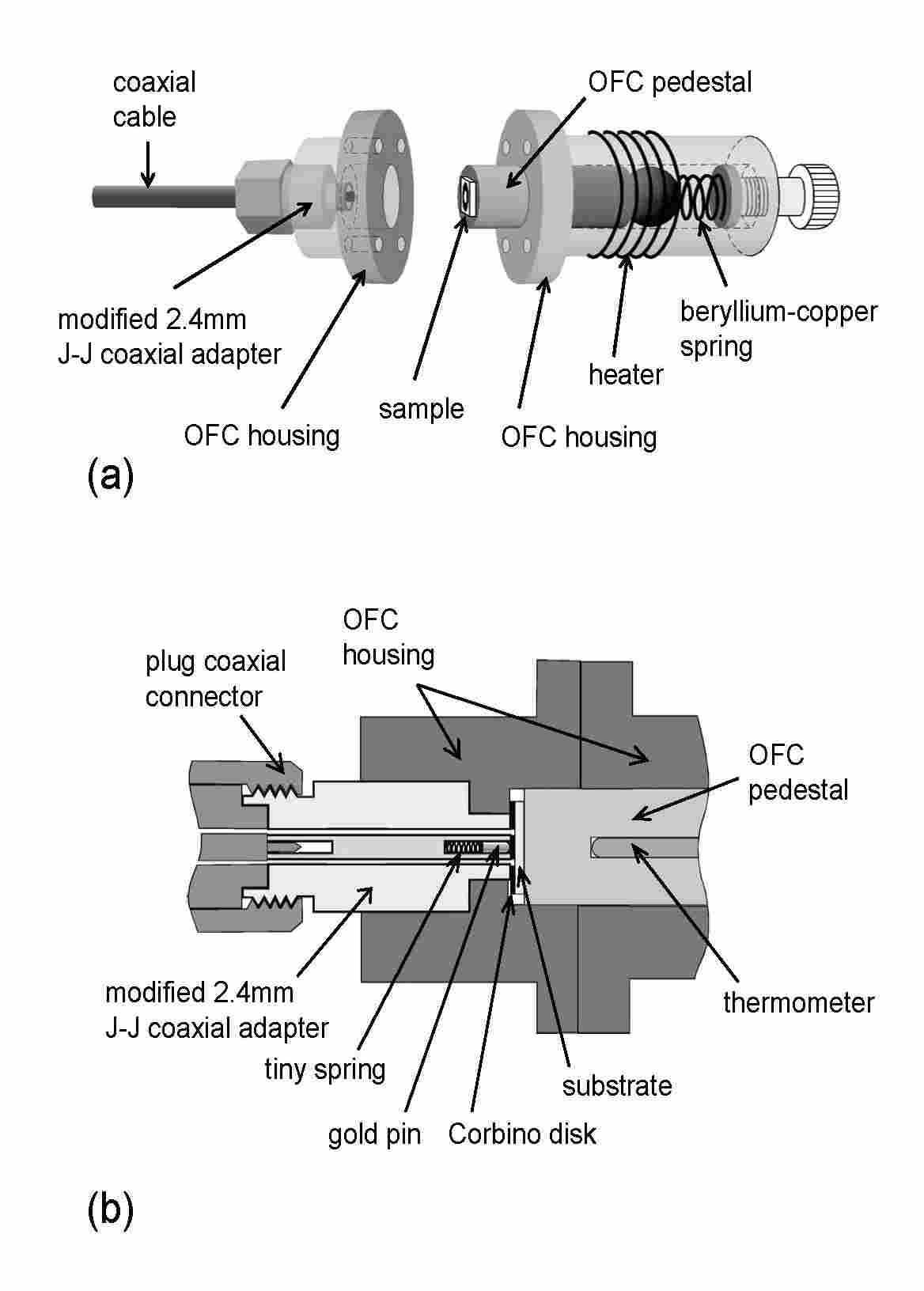}
\caption{
(a) Schematic view of coaxial microwave probe, which is located at the end of the coaxial assembly in the low temperature apparatus. 
A thin film sample is directly contacted to a modified 2.4~mm jack-to-jack adapter, by using a cylindrical oxygen-free-copper (OFC) pedestal. 
(b) Cross-sectional view of coaxial microwave probe. 
Stable electrical contact at low temperatures is realized by a tiny spring embedded into the center conductor of the modified adapter and a beryllium-copper spring in the backside of the OFC pedestal. 
A thermometer to measure the sample temperature is also inserted into the OFC pedestal. 
}
\label{fig:probe}
\end{figure}

The details of a coaxial microwave probe, which is located at the end of the coaxial assembly in the low temperature apparatus, is illustrated in Figs.~\ref{fig:probe}(a) and \ref{fig:probe}(b). 
In our apparatus, a thin film sample is connected to the coaxial line through a modified 2.4 mm jack-to-jack coaxial adapter. 
As shown in Fig.~\ref{fig:probe}(b), one side of the adapter jack was manufactured to realize a direct contact with a thin film sample, leaving only a part of the outer conductor of the 2.4 mm female connector. 
In addition, a tiny spring \cite{small_spring} and a gold pin (0.4~mm diameter and 1.6~mm length) were put into the center conductor of the female connector in the manufactured side of the adapter. 
Such a spring-loaded center conductor pin played an essential role to maintain a stable electrical contact even at low temperatures, together with a beryllium-copper spring sustaining the sample from the backside. 
The modified adapter is carefully attached to a sample housing made of oxygen free copper (OFC) so that the top of the manufactured outer conductor is adjusted to the same position as the bottom of the sample housing. 

The thin film sample with the contact electrodes in a Corbino disk geometry is placed in a space enclosed by the sample housing and a cylindrical OFC pedestal with a diameter of 7~mm. 
The inner and outer diameters of the Corbino disk are 1.0~mm and 2.4~mm, respectively, which are the same as those of the 2.4~mm coaxial connector. 
The electric contacts are typically made by painting ink with gold nano-particles \cite{perfectgold} or a standard gold (or silver) evaporation with the thickness of a few hundreds of nanometers through a washer-shaped shadow mask \cite{BoothPhD}. 
Since the microwave broadband method is in principle a two-probed measurement, we need to consider the influence of the contact resistance. 
It is particularly important for a highly conductive sample, since the lower frequency part of the magnitude of the reflected signal is seriously affected by the contact resistance, as discussed in detail in Appendix B. 
Thus, the various conditions including the choice of materials for electrodes, that of methods to make electrodes, and details of annealing treatments were experimentally determined for each sample, in order to minimize the contact resistance. 

As previously reported by Booth {\it et al.} \cite{Booth1994}, this microwave broadband technique enables to measure the two-terminal dc resistance of the sample through the bias port prepared in the $S$-parameter test set. 
Indeed, we made such a simultaneous dc measurement to estimate the contact resistance or to check the stability of the electric contact at low temperatures. 
However, near the critical temperature of the superconducting material, the simultaneous dc measurement should be carefully avoided, since an excess dc current can easily affect the intrinsic property attributed to the superconducting fluctuations. 

The stailess-steel coaxial cable is weakly anchored to the helium bath.
A typical base temperature of the low temperature apparatus described here is below 10~K at an ambient pressure of liquid helium ($\sim$ 3~K by pumping on the helium bath). 
Although this temperature is larger than the base temperature of similar equipments specialized in the low temperature measurements below 10~K \cite{MLeePRL2001,Scheffler2005}, 
the weak thermal anchor to the helium bath enables to allow the broadband measurements from $T_c$ up to room temperature in the same run.
In addition, our apparatus fits inside a bore with 2-in. diameter of a NbTi superconducting magnet, which allows the measurements under the dc magnetic field up to 8~T, as shown in Fig.~\ref{fig:scheme}. 

All measurements were performed after keeping the coaxial microwave probe  in the cryostat at least for 10 hours, in order to ensure the equilibrium of temperature distribution in the transmission line. 
This is crucially important to obtain the high reproducibility between measurement runs, as discussed in detail in the next section.

\section{\label{sec:calib}Standard calibration method}

In practice, the frequency dependence of the complex reflection coefficient is easily affected by the changes of the attenuation and phase shifts in the intervening coaxial cable, giving rise to large systematic errors at lower temperatures and higher frequencies. In the standard error model of microwave network analysis \cite{Pozar}, the measured reflection coefficient, $S_{\rm 11}^{\rm meas}(\omega)$, is related to the actual reflection coefficient, $S_{\rm 11}(\omega)$, by the following equation: 
\begin{equation}
S_{11}^{\rm meas}(\omega)=E_D(\omega)+\frac{E_R(\omega)S_{11}(\omega)}{1-E_S(\omega)S_{11}(\omega)}, 
\label{eq:error}
\end{equation}
where, $E_D(\omega)$, $E_R(\omega)$, and $E_S(\omega)$ are complex error coefficients, representing the directivity, the reflection tracking, and the source mismatching, respectively. 
One of methods to determine these unknown error coefficients is to measure three known reference samples as a function of frequency at each temperature. 

At room temperature, the whole coaxial lines except for the modified adapter can be calibrated, since the calibration standards with the 2.4~mm coaxial connectors are commercially available \cite{CalKit}. 
However, in our broadband microwave measurements, we need to calibrate the contact plane where a thin film sample is directly connected with the spring-loaded microwave probe. 
Unfortunately, the commercial calibration standards cannot be connected at this plane. 
In addition, at low temperatures, they are no longer certified, since the microwave properties of them are not characterized at low temperatures. 
Thus, we prepared the following three reference samples which could be characterized even at low temperatures. 

\subsection{Open standard}
First of all, as an open standard, we used a free-standing teflon disk with 7~mm diameter and 1.5~mm thick. 
In the open-ended coaxial probe to investigate dielectric materials close to an ideal open (that is, $S_{\rm 11}=1$), it is important to determine a finite capacitance, $C_0$, which provides a small deviation from the ideal open through the finite impedance of the open standard ($Z_L^{\rm open}=1/i\omega\epsilon C_0$, where $\epsilon$ is the relative permittivity of open standard) 
\cite{Jiang1993,Martens2000}. 
Although it is usually a very difficult task, the influence of $C_0$ was 
found to be rather small in our load-ended probe to measure highly conductive materials. 
Thus, we roughly estimated $C_0$, based on the assumption that $C_0$ was simply given by the fringe field excited by the TEM mode outside the coaxial probe \cite{memo1}. 
A numerical calculation of $C_0$ assuming the infinite thickness of the dielectric medium with $\epsilon=1$ suggested that $C_0\sim$ 25~fF \cite{memo2}. 

Another rough estimate of $C_0$ was obtained by the room-temperature measurement of $S_{\rm 11}^{\rm meas}(\omega)$ for several dielectric substrates (teflon, quartz, and MgO) with the known value of $\epsilon$ \cite{memo3}. 
As far as such room-temperature measurements are concerned, we applied the calibration data for the unmodified 2.4~mm coaxial adapter to the modified adapter, 
since the difference of $S_{\rm 11}^{\rm meas}(\omega)$ between both adapters were very small at low frequencies. 
When $Z_c\ll |Z_L^{\rm open}|=1/\omega\epsilon C_0$, the phase part $\phi(\omega)$ of $S_{\rm 11}(\omega)$ can be fitted with a line of $\phi(\omega)=\arctan(-2\omega\epsilon C_0Z_c)$ in a good approximation. 
The fitting results of $\phi(\omega)$ at low frequencies suggested that $C_0\sim$ 12~fF, which was comparable to the above numerical estimate. 
Such a small value of $C_0$ confirms that the used teflon sample can safely be regarded as the open standard for our purpose, since the corresponding impedance of teflon at $\sim$1~GHz is larger than two orders of magnitude of $Z_c$. 

In the previous work \cite{KitanoPhysicaC}, we used a teflon rod with 20 mm long rather than the present teflon disk, to approximate an infinite thickness. 
However, we found that the measured data for the teflon disk was more reproducible than that for the long teflon rod in the low-temperature measurements. 
The systematic experiments using the teflon samples with various thickness suggested that the finite-thickness effect became prominent only for thickness less than 0.5~mm. 
Thus, we adopted the teflon disk with 1.5~mm thick as the open standard. 

\subsection{Load standard}
As a load standard, we used a NiCr film with 4~mm square, which was fabricated on a quartz substrate by thermal evaporation of a NiCr wire. 
The same contact electrodes in the Corbino disk geometry as the sample under test was made to reduce the contact resistance. 
Since the resistivity of NiCr is almost independent of temperature, 
the temperature-dependent change in the intervening coaxial cable can be more precisely calibrated by the NiCr load standard than other load standards made of pure metals \cite{Stutzman2000}. 

The thickness of the NiCr film was designed to be typically between 30~nm and 120~nm, corresponding to the resistance between 4~$\Omega$ and 12~$\Omega$ in the Corbino geometry. 
In general, it is expected that the resistance of the load standard is close to be $Z_c$ (that is, $S_{\rm 11}=0$) in order to keep the large dynamic range of broadband measurements. 
However, we rather used the NiCr film with a lower resistance, which was comparable to the resistance of our thin film samples, since we cannot neglect the influence of the dielectric substrate for a thinner NiCr film with the resistance close to 50~$\Omega$ (less than 10~nm thick), as discussed in Appendix A. 

In our calibration, $Z_L^{\rm load}$ of the load standard was regarded as a purely real, by neglecting a small parasitic inductance of the load standard. 
This is allowed for our apparatus where the free-standing load standard can be directly connected with the microwave probe, showing a sharp contrast to the previous work by Stutzman {\it et al,} where a small amount of indium solder giving rise to a slight parasitic inductance is needed to obtain the mechanical and electrical contacts \cite{Stutzman2000}. 
\begin{figure}[h]
\includegraphics[height=10cm,width=8cm]{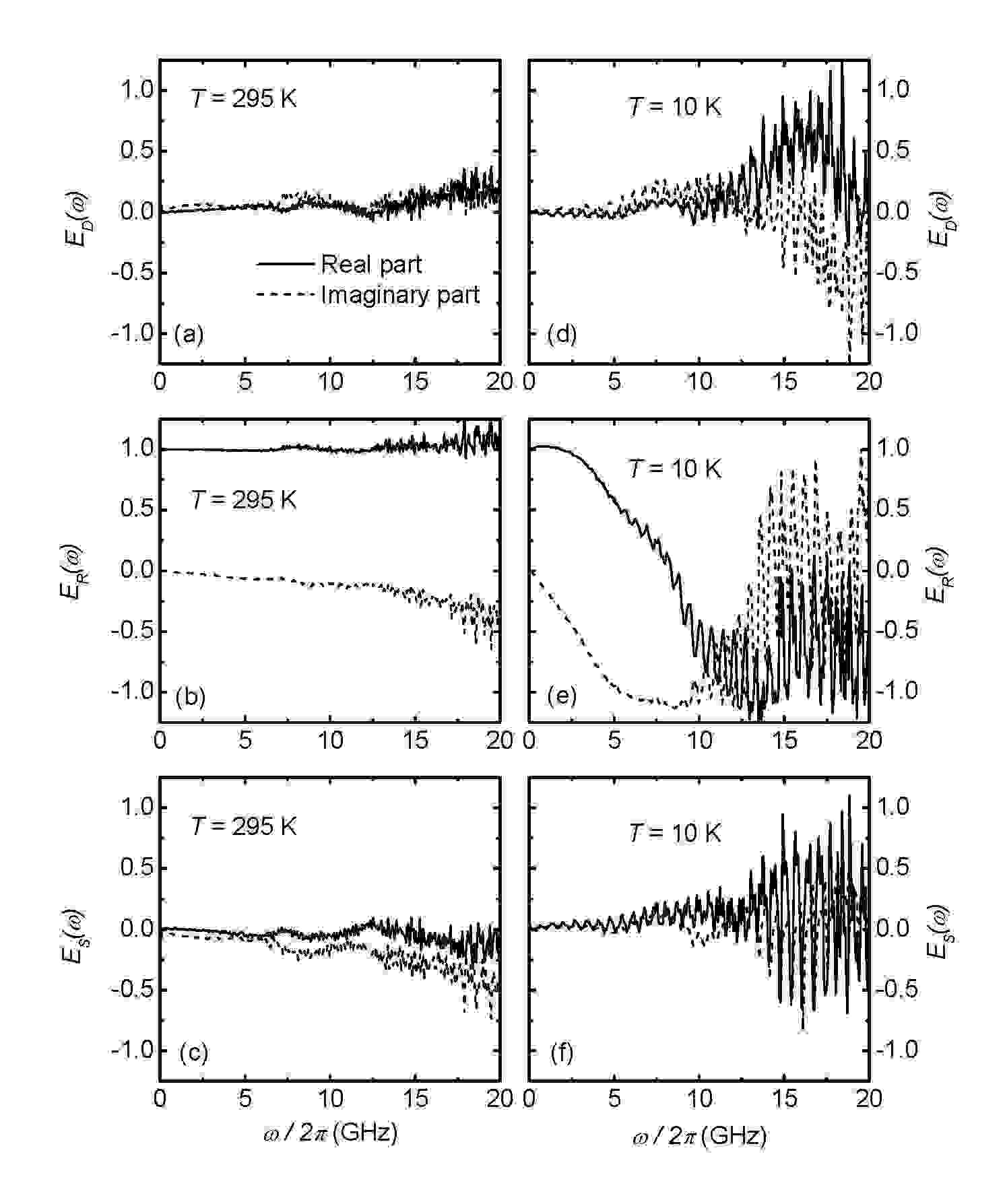}
\caption{
Typical results of the complex error coefficients, $E_D(\omega)$, 
$E_R(\omega)$, $E_S(\omega)$ at $T$=295~K (left panels) and 10~K 
(right panels). Solid and broken lines represent the real and 
imaginary part of them, respectively. 
}
\label{fig:e_coef}
\end{figure}

\subsection{Short standard}
As a short standard, we used a gold (or silver) thin film with 150~nm $\sim$ 350~nm thick, which was evaporated on a quartz substrate with 4~mm square, 
in contrast to the previous works using a piece of free-standing pure metal (for instance, aluminum or copper) \cite{Stutzman2000,Scheffler2005}. 
This is based on our experimental result that the frequency dependence of $S_{\rm 11}^{\rm meas}(\omega)$ for the aluminum plate with 0.2~mm thick was clearly stronger than that of $S_{\rm 11}^{\rm meas}(\omega)$ for the gold thin film, suggesting that the Al bulk sample could not be regarded as an ideal short with infinite conductivity (that is, $S_{\rm 11}=-1$). 

If we use a bulk of pure metal with the finite conductivity as the short standard, we must correct the influence of the frequency-dependent $Z_L^{\rm short}(\omega)(\approx\Gamma Z_S(\omega)\propto\sqrt{\omega}$, for $t\gg 1/k$), as suggested by Eq.~(\ref{eq:Zs_eff}). 
On the other hand, $Z_L^{\rm short}(\omega)(\approx 1/\sigma t)$ for a thin film sample with the finite conductivity is expected to show no frequency dependence, since $\sigma(\omega)$ can be regarded as purely real and $\omega$-independent in the measured microwave frequency region, as expected for the Drude conductivity in the so-called ``Hagen-Rubens" limit \cite{memo4}. 
Thus, we conclude that the gold (or silver) thin film evaporated on substrate is more appropriate for the short standard than the free-standing Al (or Cu) sample, as far as the influence of the substrate is negligible. 

We also made the contact electrodes with the Corbino geometry on the short standard by using the technique performed for the thin film sample under test, in order to keep the same electrical environment as the sample or load standard. 
The residual resistance of the short standard (less than 50~m$\Omega$) was experimentally estimated by the two-terminal dc resistance of the superconducting film sample at the lowest temperature. 
\\

\subsection{Determination of error coefficients}
If the complex reflection coefficients of the above-mentioned three standards (referred by $i$=1, 2, 3) were measured at each temperature as a function of frequency, one can determine $E_D(\omega)$, $E_R(\omega)$, and $E_S(\omega)$, as follows; 
\begin{widetext}
\begin{subequations}
\label{eq:EdErEs}
\begin{eqnarray}
E_D(\omega)&=&\frac{M_1(M_2-M_3)A_2A_3+M_2(M_3-M_1)A_3A_1+M_3(M_1-M_2)A_1A_2}
{(M_1-M_2)A_1A_2+(M_2-M_3)A_2A_3+(M_3-M_1)A_3A_1},\\
E_R(\omega)&=&\frac{(M_1-M_2)(M_2-M_3)(M_3-M_1)(A_1-A_2)(A_2-A_3)(A_3-A_1)}
{[(M_1-M_2)A_1A_2+(M_2-M_3)A_2A_3+(M_3-M_1)A_3A_1]^2},\\
E_S(\omega)&=&\frac{M_1(A_2-A_3)+M_2(A_3-A_1)+M_3(A_1-A_2)}
{(M_1-M_2)A_1A_2+(M_2-M_3)A_2A_3+(M_3-M_1)A_3A_1},
\end{eqnarray}
\end{subequations}
\end{widetext}
where $M_i$ and $A_i$ ($i$=1, 2, 3) are $S_{\rm 11}^{\rm meas}(\omega)$ and $S_{\rm 11}(\omega)$ for the $i$-th standard, respectively. 
Note that the expressions of the error coefficients given here are the same as those derived by Scheffler and Dressel \cite{Scheffler2005}. 

Figure \ref{fig:e_coef} shows the typical results of the real and imaginary parts of $E_D(\omega)$, $E_R(\omega)$, and $E_S(\omega)$ at both $T$=295~K and $T$=10~K. 
At room temperature ($T$=295~K), we confirmed that the frequency dependences of $E_D(\omega)$, $E_R(\omega)$, and $E_S(\omega)$ were very weak up to $\sim$15~GHz, and that only the real part of $E_R(\omega)$ was close to unity while other components of $E_D(\omega)$, $E_R(\omega)$, and $E_S(\omega)$ were almost zero. 
Such results at room temperature clearly indicate that the difference between $S_{\rm 11}^{\rm meas}(\omega)$ and $S_{\rm 11}(\omega)$ described by Eq.~(\ref{eq:error}) is negligibly small up to $\sim$15~GHz. 
On the other hand, at a low temperature ($T$=10~K), both the real and imaginary parts of $E_R(\omega)$ showed a strong frequency dependence, which was attributed to the changes of the attenuation and phase-shift in the coaxial cable with decreasing temperature. 
The frequency dependences of $E_D(\omega)$ and $E_S(\omega)$ also became more prominent above $\sim$15~GHz than those at room temperature. 
These results strongly suggest the importance of the full calibration using $E_D(\omega)$, $E_R(\omega)$, and $E_S(\omega)$ for the precise measurement at low temperatures. 
\begin{figure}[t]
\includegraphics[height=10.0cm,width=7.5cm]{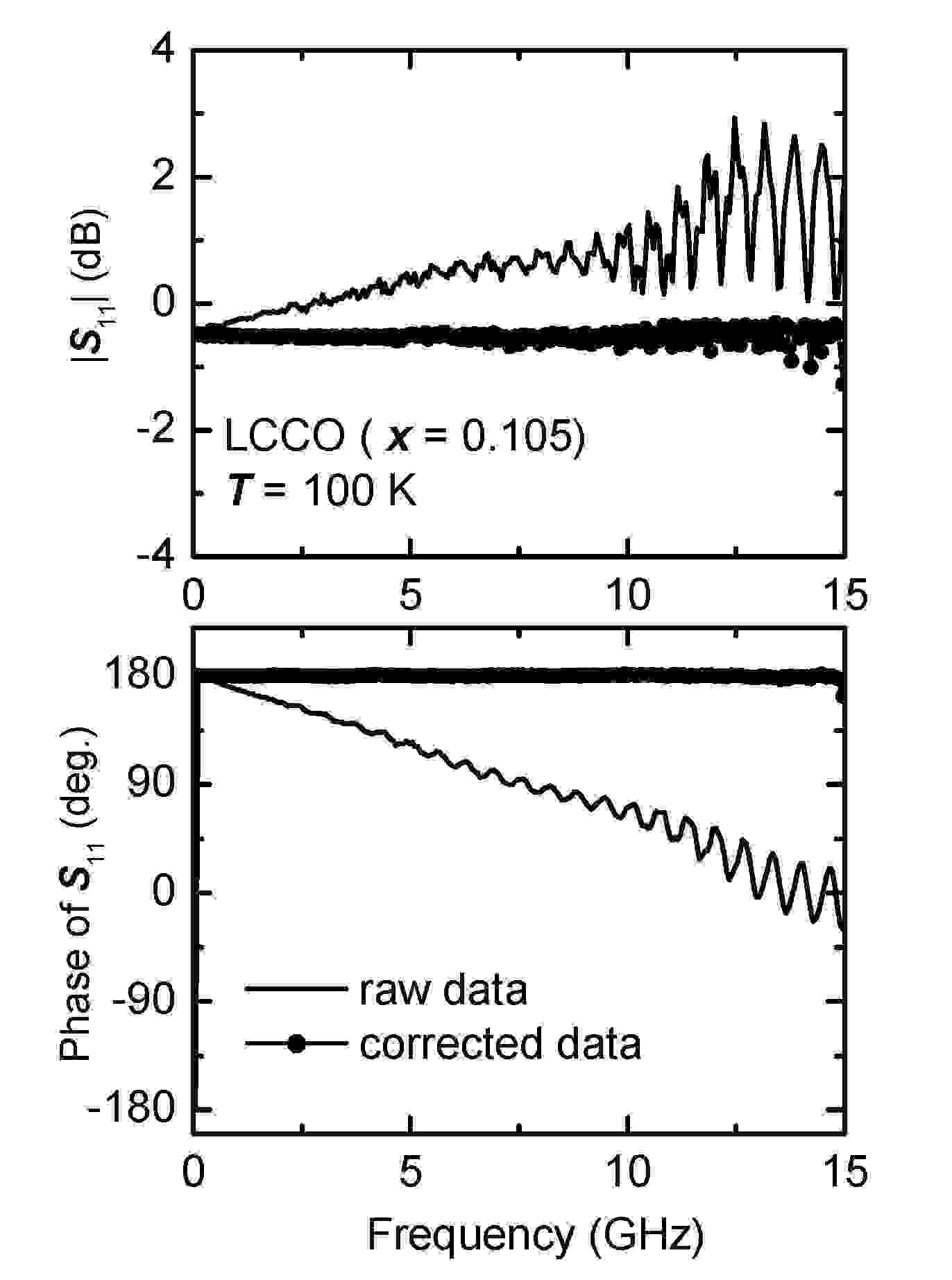}
\caption{
Magnitude (upper panel) and phase (lower panel) parts of $S_{\rm 11}(\omega)$ 
for a La$_{2-x}$Ce$_x$CuO$_4$ ($x$=0.105) thin film measured at $T$=100~K. 
Solid thick lines are the raw data before the calibration and solid circles with thin line are the corrected data by the standard calibration method. 
}
\label{fig:lcco_S11}
\end{figure}

By applying the three error coefficients determined by Eq.~(\ref{eq:EdErEs}) to Eq.~(\ref{eq:error}), we can obtain the actual data of $S_{\rm 11}(\omega)$ for unknown samples. 
Figure \ref{fig:lcco_S11} shows the typical results at $T$=100~K of the magnitude ($|S_{\rm 11}|$) and the phase ($\angle S_{\rm 11}$) parts of $S_{\rm 11}(\omega)$ for a La$_{2-x}$Ce$_x$CuO$_4$ (LCCO) thin film with $x$=0.105 and $t$=90~nm, 
which was grown on LaSrAlO$_4$ (001) substrate by molecular beam epitaxy (MBE) method \cite{Naito2000}. 
Similar results for La$_{2-x}$Sr$_x$CuO$_4$ (LSCO) thin films were already reported by the previous work \cite{KitanoPhysicaC}. 
As clearly shown in Fig.~\ref{fig:lcco_S11}, 
the ripple structure of the interference introduced by the coaxial line and the spurious frequency dependence of $S_{\rm 11}^{\rm meas}(\omega)$ due to the decrease of attenuation and cable-length were almost completely removed from the raw data. 
The measurement errors of $|S_{\rm 11}(\omega)|$ and $\angle S_{\rm 11}(\omega)$ in our measurement apparatus were estimated as 0.05~dB and 0.2$^\circ$ at 1~GHz, respectively, from the size of the ripple structure remained after the calibration. 
\begin{figure}[h]
\includegraphics[height=6.1cm,width=8cm]{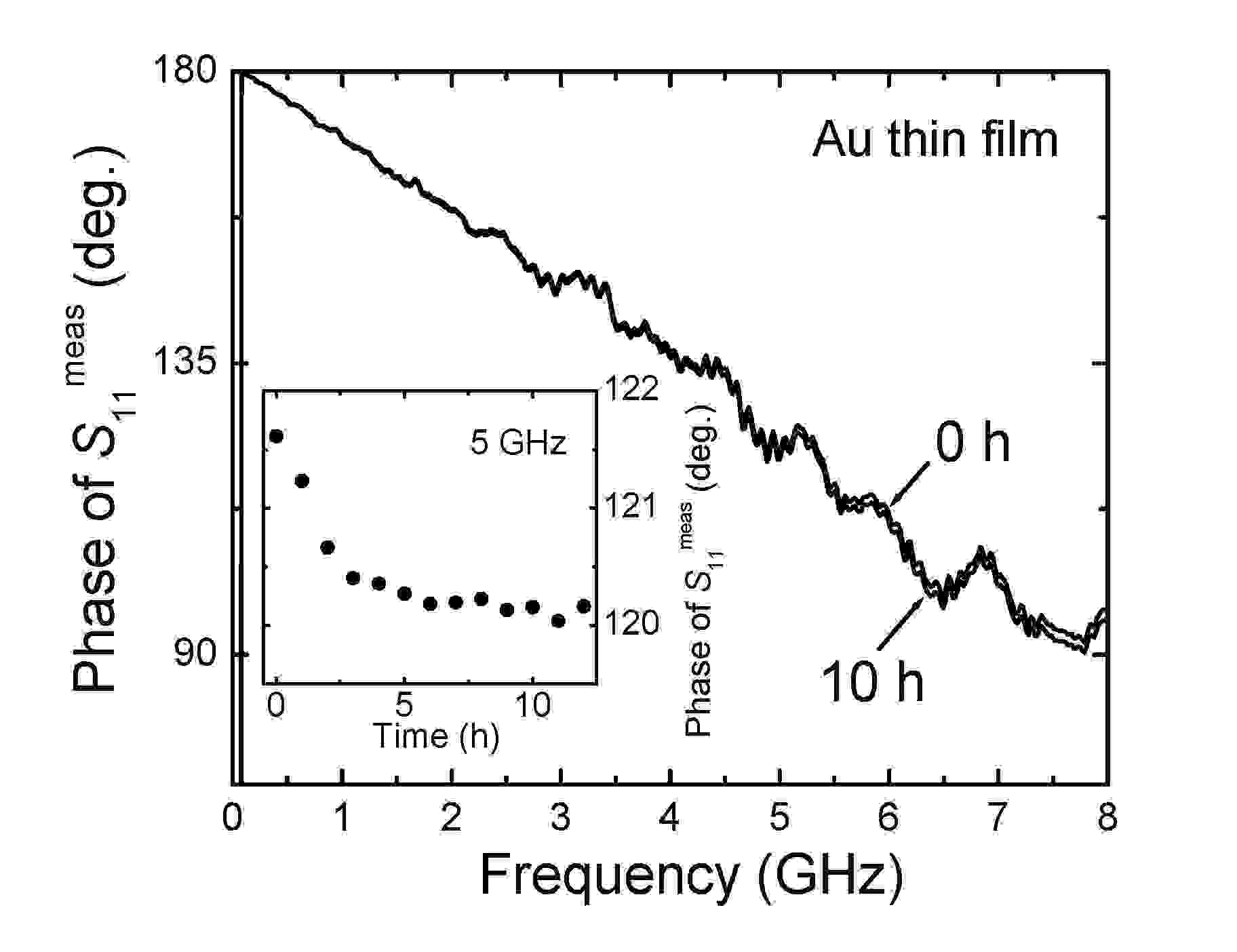}
\caption{
Frequency dependence of $\angle S_{\rm 11}^{\rm meas}(\omega)$ for a Au thin film at $T$=10~K. 
A time-dependent behavior was observed at higher frequencies than about 5~GHz. 
The inset is the time dependence of $\angle S_{\rm 11}^{\rm meas}$ at 5~GHz. 
}
\label{fig:time_dep}
\end{figure}

\subsection{Reproducibility}
As is well known, it is only systematic errors that we can correct in the calibration procedure. 
In order to reduce other random errors, the high reproducibility between four measurements including the three ({\it i.e.}, open, load, and short) standard samples and an unknown sample is crucially important. 
Thus, all the microwave coaxial connectors should be carefully fastened by using a torque wrench. 
In addition, we found that the thermal contraction of the transmission line should be taken care to achieve the high reproducibility. 

In our measurement system, the long coaxial cable ($\sim$1.25~m) included in the low temperature apparatus is roughly expected to contract at a rate of 20~$\mu$m per 1~K. 
This length corresponds to $\sim$0.05~$^\circ$/GHz in the slope of $\angle S_{\rm 11}^{\rm meas}(\omega)$ at lower frequencies. 
Usually, the thermal contraction of the dielectric spacer in the coaxial cable is much slower than that of the metallic outer and center conductors in the coaxial cable, giving rise to the time-dependent relaxation effect. 
Indeed, we observed such a time-dependent behavior became prominent in $\angle S_{\rm 11}^{\rm meas}(\omega)$ above 5~GHz, as shown in 
Fig.~\ref{fig:time_dep}. 
The inset of Fig.~\ref{fig:time_dep} shows that it takes about 10 hours to achieve the equilibrium of thermal contraction in the coaxial cable. 

In practice, we found that the reproducibility of $S_{\rm 11}^{\rm meas}(\omega)$ in our measurement system was within the above-mentioned measurement errors of $S_{\rm 11}^{\rm meas}(\omega)$. 
The limited reproducibility gives a maximum frequency where the complex conductivity $\sigma(\omega)$ of unknown samples can be successfully obtained by the standard calibration method. 
In our measurement system, the maximum frequency was typically 12~GHz (in some cases, 40~GHz), as described in the next subsection. 
On the other hand, we found that a minimum frequency where $\sigma(\omega,T)$ could be obtained (typically, $\sim$0.1~GHz), was determined by the residual contact resistance, as will be discussed in Appendix B. 

\subsection{Performance of the standard calibration method}
\begin{figure}[t]
\includegraphics[height=5cm,width=8cm]{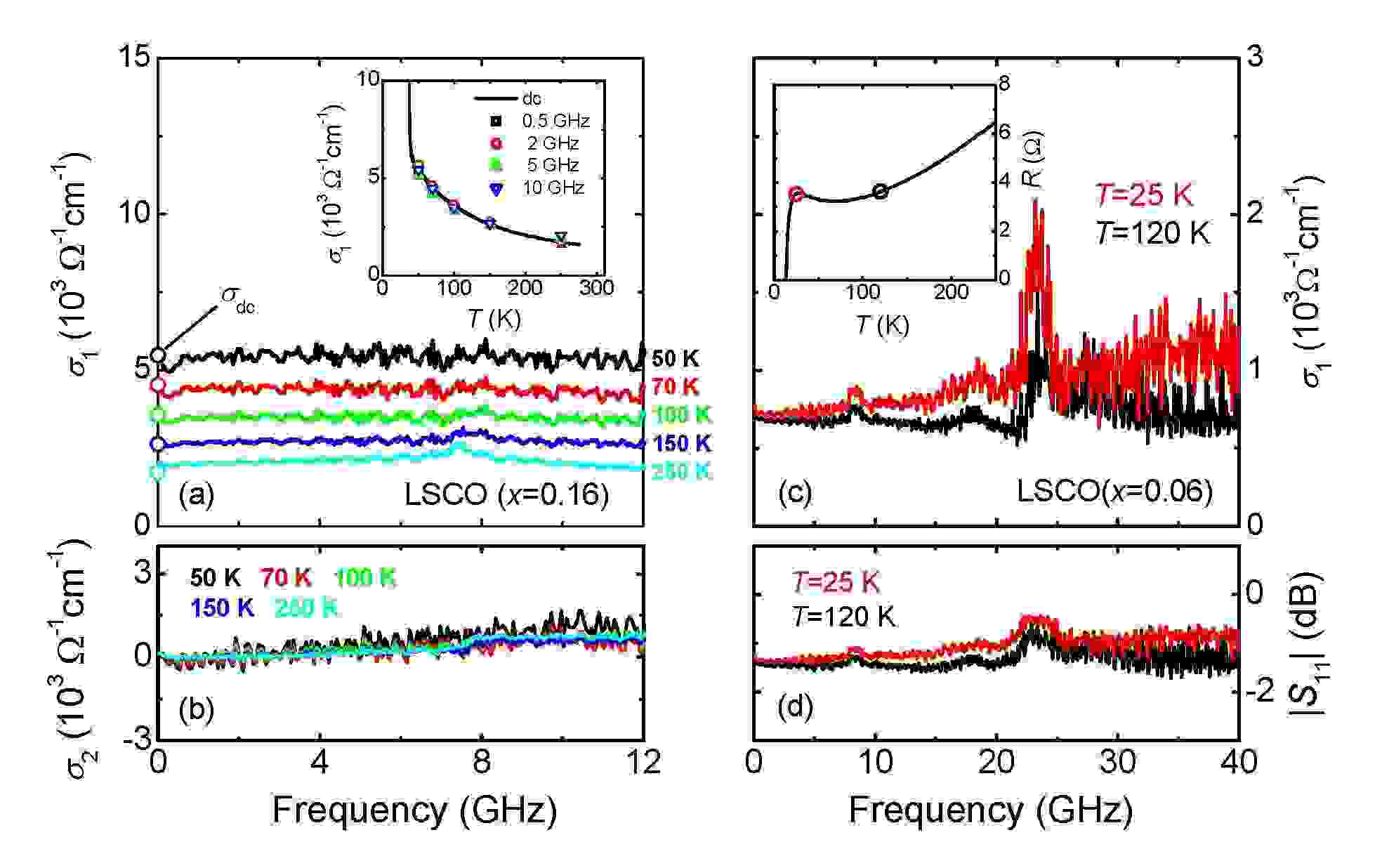}
\caption{
(a) Typical results of $\sigma_1(\omega,T)$ for a LSCO ($x$=0.16, $t$=140~nm) thin film at several temperatures far above $T_c$, which were obtained by the standard calibration method. The inset is the temperature dependence of $\sigma_1(T)$ at several frequencies together with the dc conductivity (solid line). 
(b) Frequency dependence of $\sigma_2(\omega,T)$ for the LSCO ($x$=0.16, $t$=140~nm) at the same temperatures as those used in (a). 
(c) Similar plots of $\sigma_1(\omega)$ for another LSCO ($x$=0.06, $t$=470~nm) film at $T$=25~K and 100~K. The inset is the temperature dependence of the dc resistance. 
Two open circles represent the values at the temperatures depicted in the main panel. (d) Frequency dependence of $|S_{\rm 11}(\omega)|$ for the LSCO ($x$=0.06, $t$=470~nm) thin film at $T$=25~K and 100~K. 
}
\label{fig:sgm_normal}
\end{figure}

Figures \ref{fig:sgm_normal}(a) and \ref{fig:sgm_normal}(b) show the typical results of the frequency dependence of both real and imaginary parts of $\sigma(\omega,T)$ for a LSCO ($x$=0.16, $t$=140~nm) thin film at several temperatures far above $T_c$, which were obtained by the standard calibration method described in this section. 
These results clearly suggest that we could obtain the normal-state complex conductivity for the LSCO thin film at least up to 12~GHz in our measurement system, since both $\sigma_1(\omega,T)$ and $\sigma_2(\omega,T)$ for the LSCO thin film in the normal state successfully agreed with the behaviors expected in the Hagen-Rubens limit of the Drude conductivity. 

In some cases where the four broadband measurements for the open, short, load,  and the LSCO samples were reproducible \cite{Reproducibility}, 
we could obtain $\sigma_1(\omega,T)$ for the LSCO ($x$=0.06, $t$=470~nm) sample even up to $\sim$40~GHz, except for a few resonant peaks attributed to the cavity resonance in the dielectric substrate (Details of such peaks will be discussed in Appendix A), as shown in Figs.~\ref{fig:sgm_normal}(c) and \ref{fig:sgm_normal}(d). 
Interestingly, we observed the slight positive slope in the frequency dependences of both $|S_{\rm 11}(\omega)|$ and $\sigma_1(\omega)$ at $T$=25~K, 
where $\sigma_{\rm dc}(T)$ for the same LSCO sample showed the insulating behavior, as shown in the inset of Fig.~\ref{fig:sgm_normal}(c). 
This positive slope in $\sigma_1(\omega)$ was never observed at 100~K, where the magnitude of $\sigma_{\rm dc}(T)$ was almost the same as $\sigma_{\rm dc}$ at $T$=25~K, while the temperature dependence of $\sigma_{\rm dc}(T)$ was metallic. 
A recent optical study of the lightly hole-doped LSCO ($x$=0.03 and 0.04) single crystals reported that the Drude-like behavior in the optical conductivity above 80~K where $\sigma_{\rm dc}(T)$ was metallic evolved into a peak at finite frequency around 100~cm$^{-1}$ at lower temperatures, where $\sigma_{\rm dc}(T)$ was insulating \cite{DummPRL2003}. 
Since the peak of $\sigma_1(\omega)$ evolving in the far infrared region provides a finite positive slope of $\sigma_1(\omega)$ in the microwave region, we concluded that the finite positive slope of $\sigma_1(\omega)$ observed at $T$=25~K suggested the intrinsic property to the lightly doped LSCO thin film. 

These results confirm that we can obtain $\sigma(\omega,T)$ for the metallic thin film samples up to 12~GHz (in some case, up to 40~GHz) in our measurement system by using the standard calibration method. 
However, near the superconducting transition temperature, we found that the standard calibration method was no longer useful, as will be described in the next section. 
\begin{figure}[t]
\includegraphics[height=5.3cm,width=8cm]{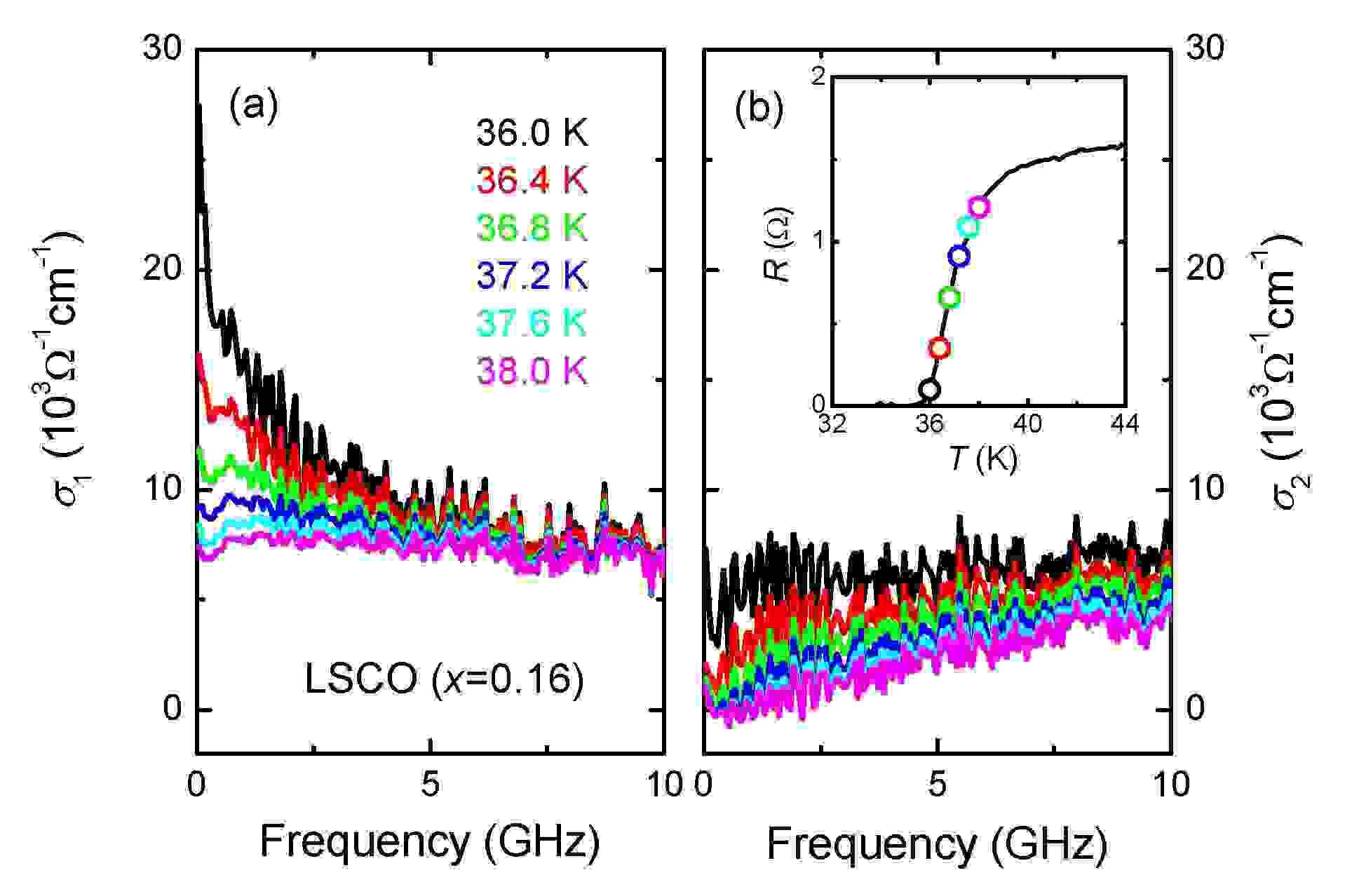}
\caption{
Frequency dependence of the real (a) and imaginary (b) parts of $\sigma(\omega,T)$ 
for the LSCO ($x$=0.16, $t$=140~nm) film near $T_c$, which were obtained by the standard calibration method. The inset is the temperature dependence of the dc resistance for the same film. Open circles are the dc resistance at the temperatures depicted in the main panel.
}
\label{fig:sgm_nearTc}
\end{figure}

\section{New calibration method for superconducting films near the critical temperature}
\subsection{Failure of the standard calibration method near $T_c$}
In our measurement system, both the real and imaginary parts of $\sigma(\omega,T)$ could be successfully obtained down to 10~K in the frequency range between 0.1~GHz and 12~GHz, by using the standard calibration method. 
However, we found that such a calibration method was insufficient to obtain $\sigma(\omega,T)$ near the superconducting transition temperature, where the strongly frequency-dependent $\sigma(\omega)$ was induced by the superconducting fluctuations. 
Figure \ref{fig:sgm_nearTc} shows the frequency dependence of $\sigma(\omega,T)$ for the LSCO ($x$=0.16) thin film at several temperatures just above $T_c$, which was obtained by the calibration method described in the previous section. 
It is clear that $\sigma_1(\omega,T)$ shows some diverging behavior in the low frequency limit as the temperature approaches toward $T_c$ from above, although $\sigma_2(\omega,T)$ shows no diverging feature even in the vicinity of $T_c$. 
This is quite unusual, since the effects of the superconducting fluctuation should affect both of $\sigma_1(\omega,T)$ and $\sigma_2(\omega,T)$ near $T_c$, according to the conventional theory of the superconducting fluctuations \cite{Schmidt}. 
\begin{figure}[t]
\includegraphics[height=9.4cm,width=8cm]{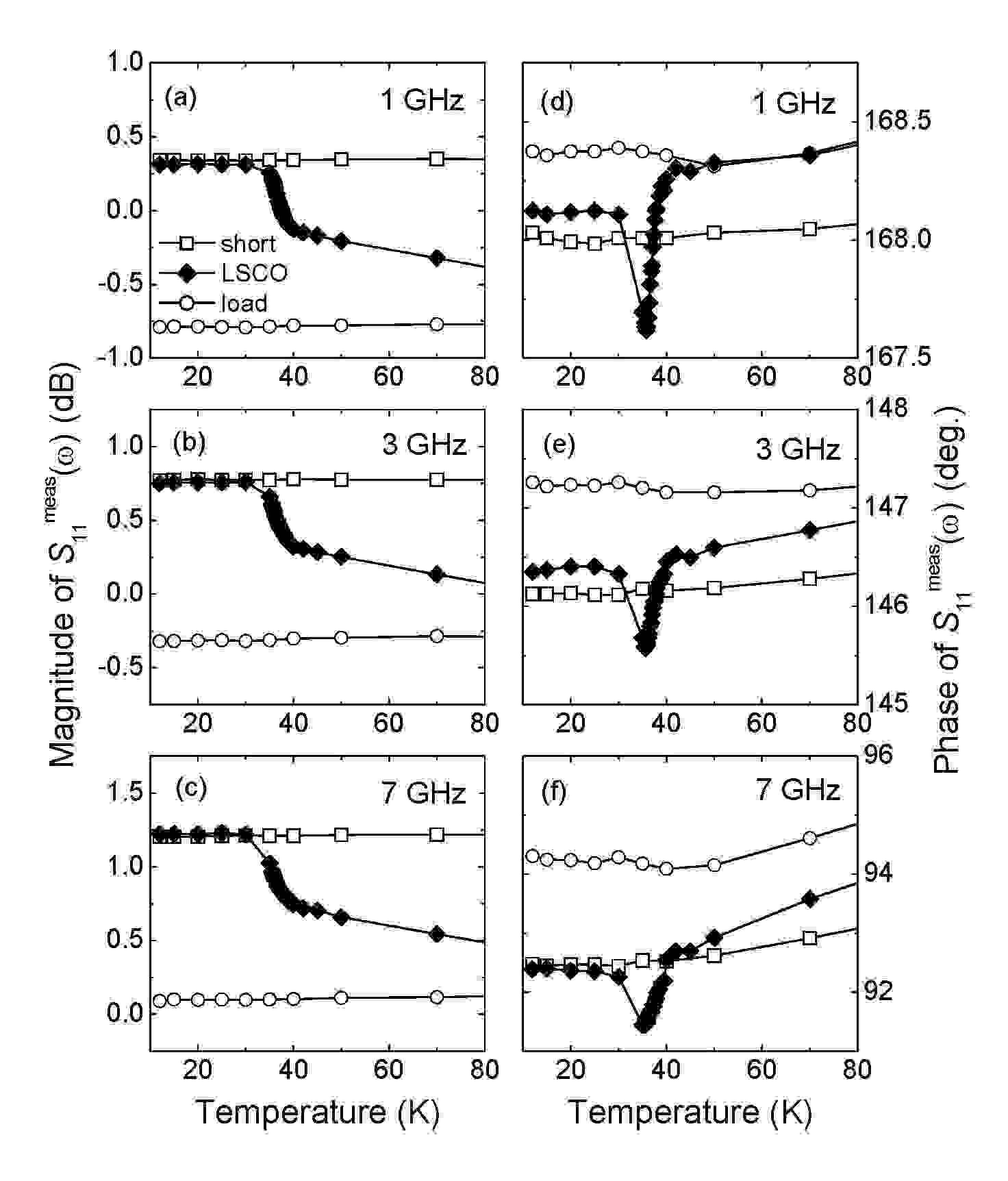}
\caption{
Temperature dependence of the magnitude (left panels) and the phase (right panels) of $S_{\rm 11}(\omega,T)^{\rm meas}$ at 1~GHz (upper panels), 3~GHz (middle panels), and 7~GHz (bottom panels) for the LSCO ($x$=0.16, $t$=140~nm) (solid diamond), a short standard (open square), and a load standard (open circle). 
}
\label{fig:S11_Tdep}
\end{figure}

As shown in Fig.~\ref{fig:S11_Tdep}, the plots of $|S_{\rm 11}^{\rm meas}(T)|$ and $\angle S_{\rm 11}^{\rm meas}(T)$ at the fixed frequency suggested the characteristic change due to the superconducting transition was properly measured. 
We confirmed that such a behavior was observed up to 10~GHz, suggesting that the contribution of the superconducting fluctuations to $\sigma(\omega,T)$ in the vicinity of $T_c$ could be measured by 
the present apparatus at least up to 10~GHz. 
We found that the failure was rather attributed to a small difference between $\angle S_{\rm 11}^{\rm  short}(\omega)$ and $\angle S_{\rm 11}^{\rm load}(\omega)$ (typically $0.2~^\circ\sim0.3~^\circ$ at 1~GHz). 
Note that there is no difference between $\angle S_{\rm 11}^{\rm  short}(\omega)$ and $\angle S_{\rm 11}^{\rm load}(\omega)$ if the broadband measurements are completely reproducible. 
Indeed, the observed difference between $\angle S_{\rm 11}^{\rm  short}(\omega)$ and $\angle S_{\rm 11}^{\rm load}(\omega)$ was comparable to the measurement errors of $\angle S_{\rm 11}^{\rm meas}(\omega)$ at the same frequency, 
suggesting that it was not a serious problem except for the temperature region near $T_c$. 
However, Eq.~(\ref{eq:sigma}) suggests that a small error in $S_{\rm 11}(\omega)$ gives rise to a large error in $\sigma(\omega)$ as $S_{\rm 11}$ becomes close to -1 in the vicinity of $T_c$. 
Thus, the standard calibration method is no longer useful to measure $\sigma(\omega,T)$ for the superconducting film  near $T_c$.

\subsection{New calibration method for a superconducting film}

In order to overcome this difficulty, we modified the calibration method as follows. 
First of all, we make sure that $\sigma(\omega,T)$ of the superconducting film measured at a temperature ($T=T_0$) far above $T_c$, which is obtained through the standard calibration, can be regarded as the Drude conductivity in the Hagen-Rubens limit, as shown in Figs.~\ref{fig:sgm_normal}(a) and \ref{fig:sgm_normal}(b). 
Next, based on this confirmation, we regard both $|S_{\rm 11}^{\rm meas}(\omega,T_0)|$ and $\angle S_{\rm 11}^{\rm meas}(\omega,T_0)$ for the superconducting thin film sample as those for the load standard. 
Furthermore, we equate $\angle S_{\rm 11}^{\rm short}(\omega,T_0)$ to $\angle S_{\rm 11}^{\rm load}(\omega,T_0)$, based on the assumption that the observed difference between them is purely due to the limited reproducibility. 
\begin{figure}[t]
\includegraphics[height=5.4cm,width=8cm]{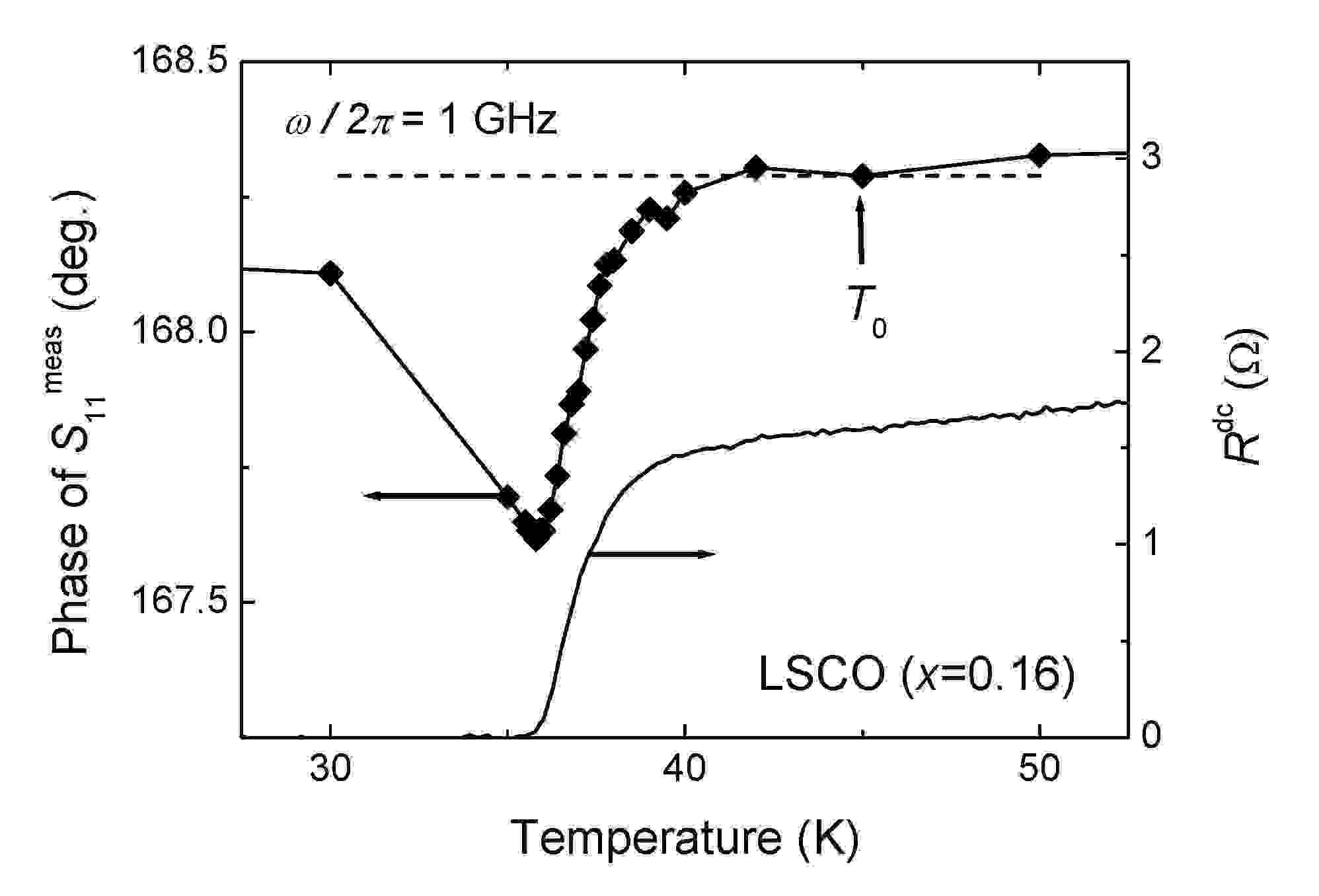}
\caption{
Temperature dependence of $\angle S_{\rm 11}^{\rm meas}(\omega,T)$ at 1~GHz and the dc resistance for the LSCO ($x$=0.16, $t$=140~nm) thin film. 
$T_0$, which is far above $T_c$, is a temperature where the data of $S_{\rm 11}^{\rm meas}(\omega)$ for the sample is regarded as a load standard. 
Broken straight line represents $\angle S_{\rm 11}^{\rm meas}(T)$ at 1~GHz for both the load and short standards used in the new calibration method. 
}
\label{fig:new_calib}
\end{figure}

Finally, we assume that both $\angle S_{\rm 11}^{\rm short}(\omega,T)$ and $\angle S_{\rm 11}^{\rm load}(\omega,T)$ were independent of temperature in a relatively narrow range between $T_c$ and $T_0$, as shown in Fig.~\ref{fig:new_calib}. 
Note that this new calibration procedure is valid only for the narrow temperature region near $T_c$, since we neglect the temperature dependence of $\angle S_{\rm 11}^{\rm short}$ and $\angle S_{\rm 11}^{\rm load}$. 
In addition, this calibration procedure cannot be applied to a case that the normal-state conductivity of a superconducting sample in the measured frequency region cannot be regarded as the Drude conductivity in the Hagen-Rubens limit, as was suggested for some heavy fermion compounds \cite{Scheffler2005}. 

In this new calibration procedure, we can reduce the measurements necessary for the full calibration. 
For example, the measurement for the load standard can be replaced by that for the superconducting sample far above $T_c$, if the normal-state conductivity of the sample can be regarded as the Drude conductivity in the Hagen-Rubens limit. 
Thus, the measurements necessary for the calibration can be reduced from four to three (sample, short and open). 
The use of the data sets in the same run also enhances the quality of the calibration. 
However, a careful consideration is needed to replace the measurement of $|S_{\rm 11}^{\rm short}(\omega)|$ by the measurement for the superconducting sample far below $T_c$, since the dissipation of the superconducting sample at a finite frequency is in principle finite at finite temperatures. 
This issue is similar to the estimates of $C_0$ in the open standard for dielectric materials, as was discussed in Sec.~IV.A. 
In practice, only in the measurements of conventional superconductors, we replaced the measurement of $|S_{\rm 11}^{\rm short}(\omega)|$ by that for the superconducting sample below $T_c$ \cite{OhashiPRB2006}. 
For high-$T_c$ cuprate superconductors, we adopted the measurements for the Au film as that for the short standard \cite{KitanoPRB2006}, since the origin of the dissipative conductivity in the superconducting state of these materials was still debated \cite{CorsonPRL2000}. 
\begin{figure}[t]
\includegraphics[height=5.6cm,width=8cm]{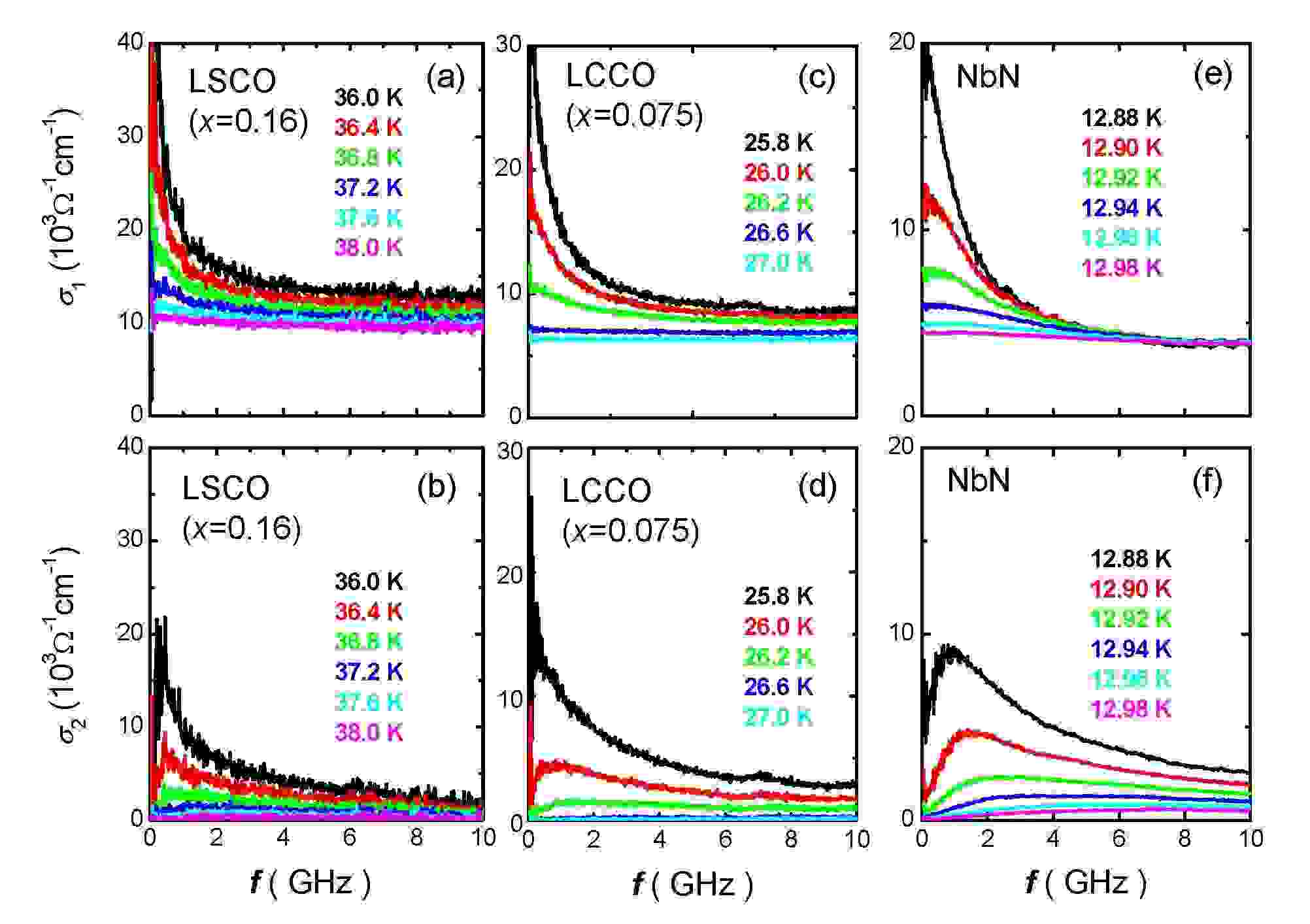}
\caption{
Frequency dependence of $\sigma_1(\omega,T)$ (upper panels) and $\sigma_2(\omega,T)$ (lower panels), which was obtained by the new calibration method, for LSCO ($x$=0.16, $t$=140~nm), LCCO ($x$=0.075, $t$=90~nm) and NbN ($t$=50~nm) thin films at several temperatures in the vicinity of $T_c$. 
Strongly frequency-dependent behavior due to the superconducting fluctuation effect was clearly observed in both $\sigma_1(\omega,T)$ and $\sigma_2(\omega,T)$ for all thin films, showing a sharp contrast to the plots shown in Fig.~\ref{fig:sgm_nearTc}. 
}
\label{fig:sgm_fluct}
\end{figure}

Although Booth {\it et al.} proposed a different calibration procedure for the superconducting sample, which was referred as the ``short only" calibration \cite{Booth1994,Booth1996}, we emphasize that the new calibration procedure proposed in this work is much better than the ``short only" calibration, 
for the following two reasons: 
First, the ``short only" calibration was based on the assumption that the change of $E_D$ and $E_S$ due to the cooled transmission line is negligible, compared with a large change of $E_R$ \cite{Booth1994}. 
However, our results strongly suggested that this assumption was not always satisfied quantitatively, as shown in Fig.~\ref{fig:e_coef}. 
Second, as was described above, the failure of the standard calibration using three standards for the superconducting sample is never resolved by the ``short only" calibration using the room-temperature data of $E_D$ and $E_S$, because the consideration on $\angle S_{\rm 11}$ for the load and short standards is essentially important to resolve this difficulty. 
\begin{figure}[t]
\includegraphics[height=8.3cm,width=8cm]{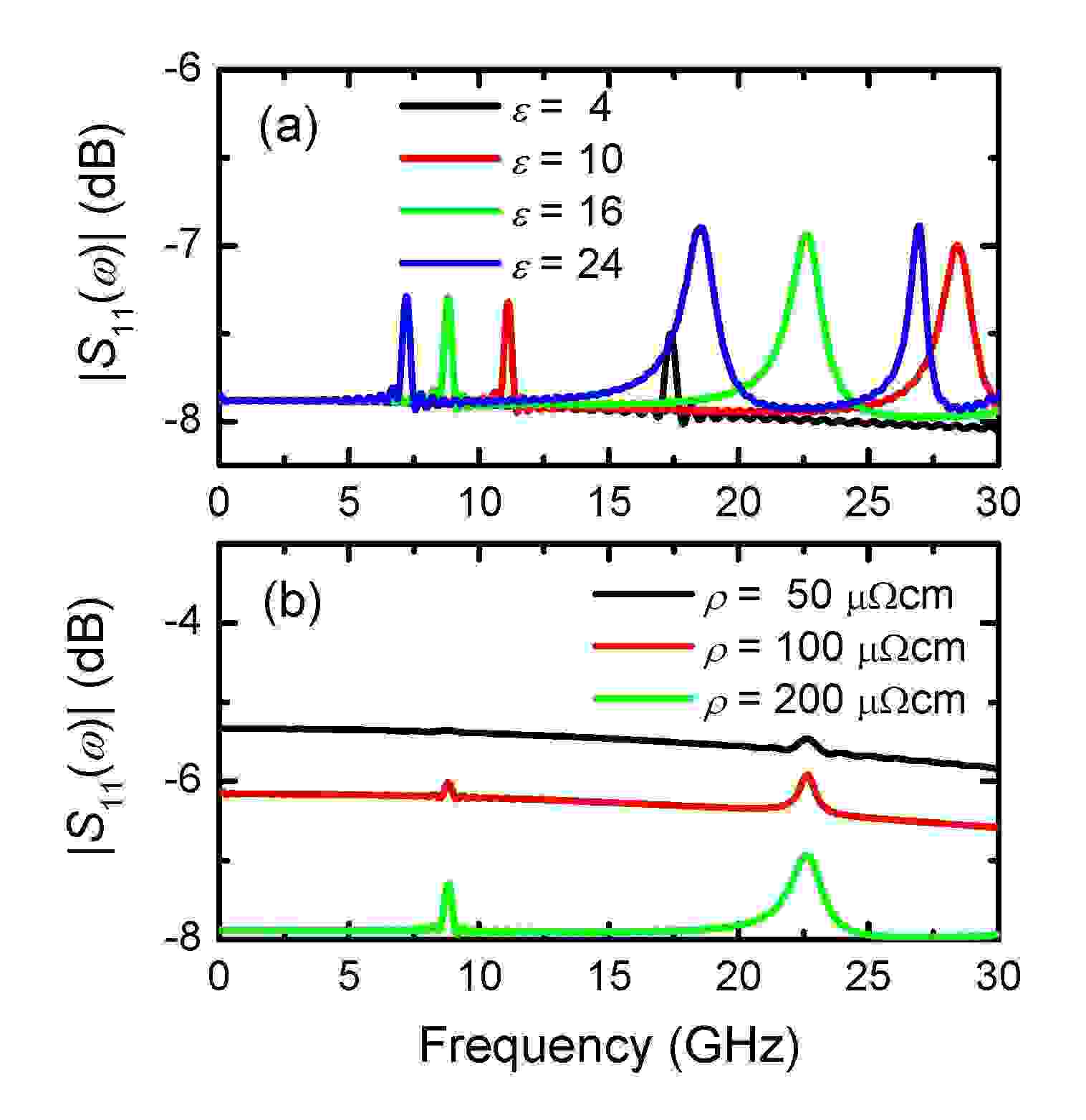}
\caption{
(a) Numerical simulations of $|S_{\rm 11}(\omega)|$ for the same boundary condition as for the actual coaxial probe, when the relative permittivity of the substrate is changed. 
(b) Similar numerical results of $|S_{\rm 11}(\omega)|$, when the resistivity of the sample film is changed. 
}
\label{fig:S11_eps_rho}
\end{figure}

Scheffler and Dressel also discussed the validity of the ``short only" calibration for a wide range of possible sample impedance and argued that the ``short only" calibration was only valid to measure highly conductive samples with $S_{\rm 11}$ close to -1 \cite{Scheffler2005}. 
We agree that the ``short only" calibration cannot be used for a sample with large impedance. 
However, our results clearly suggested that it should not be used even for a superconducting sample with $S_{\rm 11}$ very close to -1. 
Although Scheffler and Dressel presented the experimental data to compare between the ``short only" calibration and the standard calibration \cite{Scheffler2005}, the reported data do not seem to be attributed to the difference between the two calibration methods. 
Rather, it is suggested that they were related to the difference between a bulk short sample and a thin film short sample, since they did not correct the frequency dependence of $Z_L^{\rm short}(\omega)$ for the bulk short sample, which was already discussed in Sec.~IV.C.

\subsection{Performance of the new calibration method}
We present that the new calibration method described here plays a crucial role of obtaining the reliable data sets of $\sigma(\omega,T)$ for various superconducting thin films in the vicinity of $T_c$. 
Figure \ref{fig:sgm_fluct} shows the frequency dependence of $\sigma(\omega,T)$ of LSCO ($x$=0.16, $t$=140~nm), LCCO ($x$=0.075, $t$=90~nm) and NbN ($t$=50~nm) thin films at several temperatures near $T_c$. 
It is clear that both of $\sigma_1(\omega)$ and $\sigma_2(\omega)$ show a beautiful diverging feature in the low frequency limit as the temperature approaches $T_c$ from above, 
showing a sharp contrast to the behaviors appearing in Fig.~\ref{fig:sgm_nearTc}. 
We also found that this new calibration method was useful for the broadband measurement under an external magnetic field (at least, up to $B$=1~T), if the assumptions described in the previous subsection were hold \cite{memo5}. 

Details of the dynamic scaling analysis of $\sigma(\omega,T)$ obtained in the vicinity of $T_c$ strongly suggested that the new calibration method was enough reliable to determine the universality class and the dimensionality of the superconducting transition \cite{KitanoPRB2006,OhashiPRB2006}.
By applying this method to the superconducting LSCO thin films with a wide range of the Sr concentration, we succeeded in observing the anomalous change of the universality class with hole doping \cite{Ohashi_condmat}. 

Finally, the apparatus and calibration method described here have great potential to enable the exploration of the critical dynamics and the vortex dynamics for various exotic superconductors in the thin film form. 
In particular, the dynamic scaling analysis of the frequency-dependent fluctuation conductivity is rather unique, since no assumption about dimensionality and critical exponents needs to be made. 
Thus, the apparatus and calibration method presented by this work play a crucial role of the understanding of the critical charge dynamics in the superconducting transition. 
Of course, the frequency region and sensitivity realized by this work seem to be still limited for the investigation in the superconducting state far below $T_c$. 
In order to expand the measured frequency range to higher frequencies, the down-sizing of the coaxial probe and the corbino disk geometry should proceed, since it pushes the cavity resonance in the dielectric substrate to a higher frequency side. 
On the other hand, to enhance the sensitivity, the use of a shorter coaxial line or an air-line is favorable, since the background change of $\angle S_{\rm 11}(\omega)$ in the whole transmission line should be decreased to detect the slight change of $\sigma_2(\omega,T)$ in the superconducting state.  
The shortened coaxial cable also seems to be useful for the improvement of the reproducibility in the broadband measurements. 

In conclusion, we developed the microwave spectrometer to measure the complex conductivity of a superconducting thin film as a function of frequency in the temperature region from $T_c$ to room temperature. 
This spectrometer covers the frequency range between 0.1~GHz and 10~GHz even if the temperature approaches to $T_c$ from above, where $S_{\rm 11}$ is close to -1. 
By using this spectrometer, we demonstrated that $\sigma(\omega,T)$ for several thin films of both conventional and high temperature superconductors was successfully obtained even in the vicinity of $T_c$.

\begin{acknowledgments}
The authors thank H. Xu, A. Schwartz and S. Anlage for valuable discussion and comments about a broadband microwave technique, H. Akaike and A. Fujimaki for providing the NbN thin films, I. Tsukada for providing the La$_{2-x}$Sr$_x$CuO$_4$ thin films, and A. Tsukada and M. Naito for providing the La$_{2-x}$Ce$_x$CuO$_4$ thin films. This work was partly supported by the Grant-in-Aid for Scientific Research 
(13750005 and 15760003) from the Ministry of Education, Science, Sports and Culture of Japan. 

\end{acknowledgments}

\appendix

\section{Influence of a dielectric substrate}
\begin{figure}[b]
\includegraphics[height=5.8cm,width=8cm]{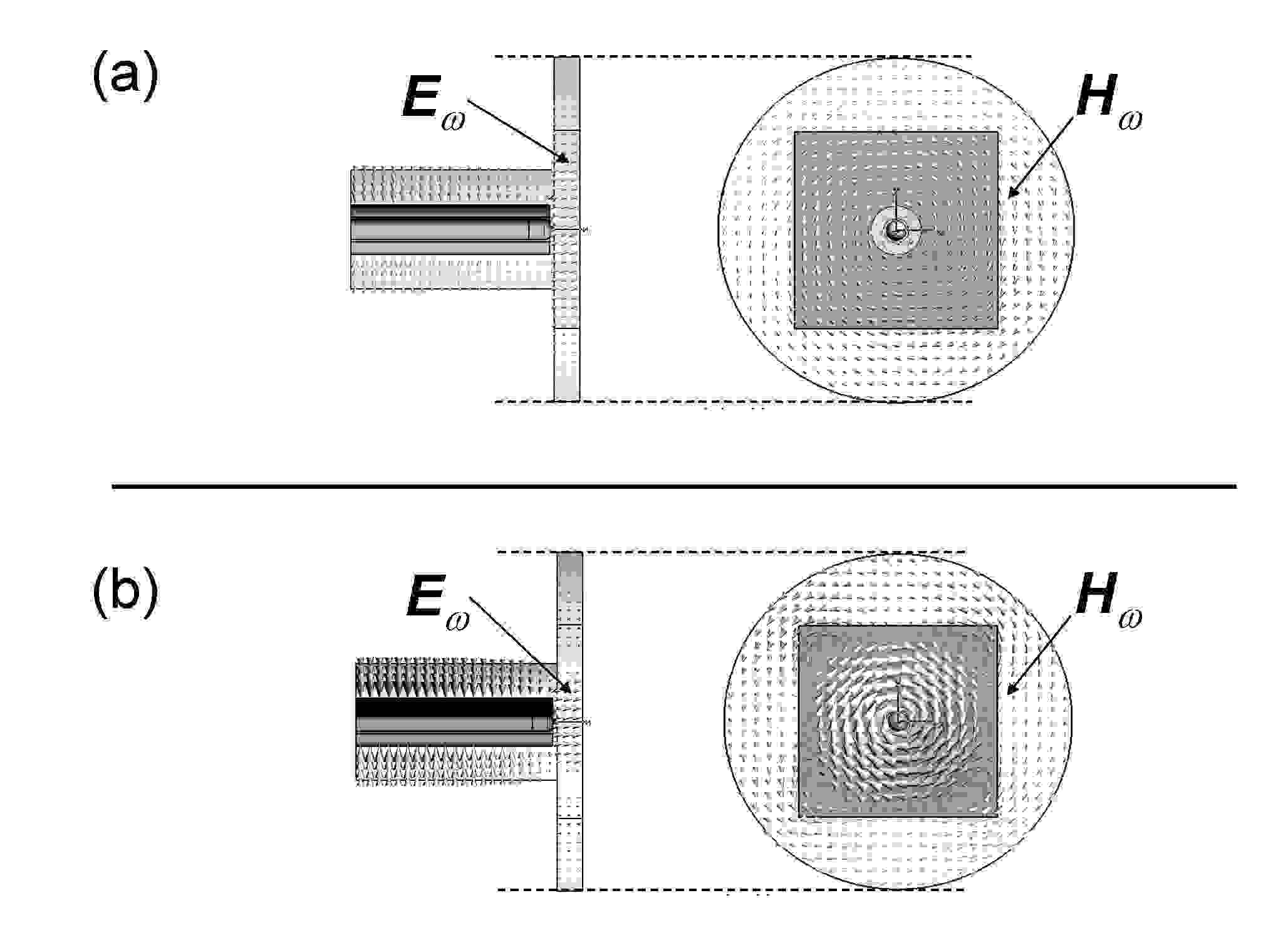}
\caption{
(a) Spatial distribution of the electromagnetic field inside  the region including a coaxial probe and a thin film sample at 8.8~GHz, corresponding to the first resonance peak when the relative permittivity of the substrate is 16 and the resistivity of the film is 200~$\mu\Omega$cm. 
(b) Spatial distribution of the electromagnetic field at 22.5~GHz, corresponding to the second resonance peak under the same condition as (a). 
Left viewgraphs are the plots of the microwave electric fields, $E_\omega$, in a cross-sectional plane including a center axis of the coaxial probe, while right viewgraphs are the plots of the microwave magnetic fields, $H_\omega$, in a plane perpendicular to the center axis, which is inside a cylindrical space (diameter is 7mm, height is 0.5 mm) including the substrate (4$\times$4$\times$0.5 mm$^3$) beneath the thin film. 
The direction and size of each cone represent the direction and magnitude of electromagnetic field at each position.  
}
\label{fig:EMfield}
\end{figure}

The effect of a dielectric substrate is an important issue to be resolved, when we measure the complex conductivity of a thin film with less conductivity and thinner thickness, since the electromagnetic field is transmitted into the dielectric substrate beneath the thin film. 
However, in practice, it is very difficult to correct such effects in a quantitatively satisfactory manner. 
The main difficulty is that the assumption of TEM-propagation, on the basis of which Eqs.~(\ref{eq:Zs_eff})$\sim$(\ref{eq:sigma}) are derived, is no longer valid in the dielectric substrate, and the contribution of TM modes should be involved appropriately, and one needs to solve the Maxwell equation on the same boundary condition as an actual experimental setup. 
Thus, the previous analyses on this issue, which assumed only the TEM-propagation \cite{Booth1994} and used an effective impedance under the boundary condition different from the actual setup \cite{Silva1996}, are too simple to resolve this problem. 

In this work, we adopted both experimental and 
numerical simulation approaches. 
Experimental results suggested that the most serious influence was an appearance of some resonance peaks, which were attributed to the fact that a substrate was sustained by a metallic pedestal in the back, as long as we used the substrate with low dielectric loss. 
As shown in Figs.~\ref{fig:sgm_normal}(c) and \ref{fig:sgm_normal}(d), we observed two resonant peaks near 8~GHz and 23~GHz for the LSCO ($x$=0.06) film. 
In the previous study \cite{KitanoPhysicaC}, we confirmed that these peaks were attributed to the use of LaSrAlO$_4$ (LSAO) substrate, since the similar peaks at the same frequencies were observed in a NiCr film on the LSAO substrate, while no peak was observed in another NiCr film on a quartz substrate. 
\begin{figure}[t]
\includegraphics[height=6.8cm,width=8cm]{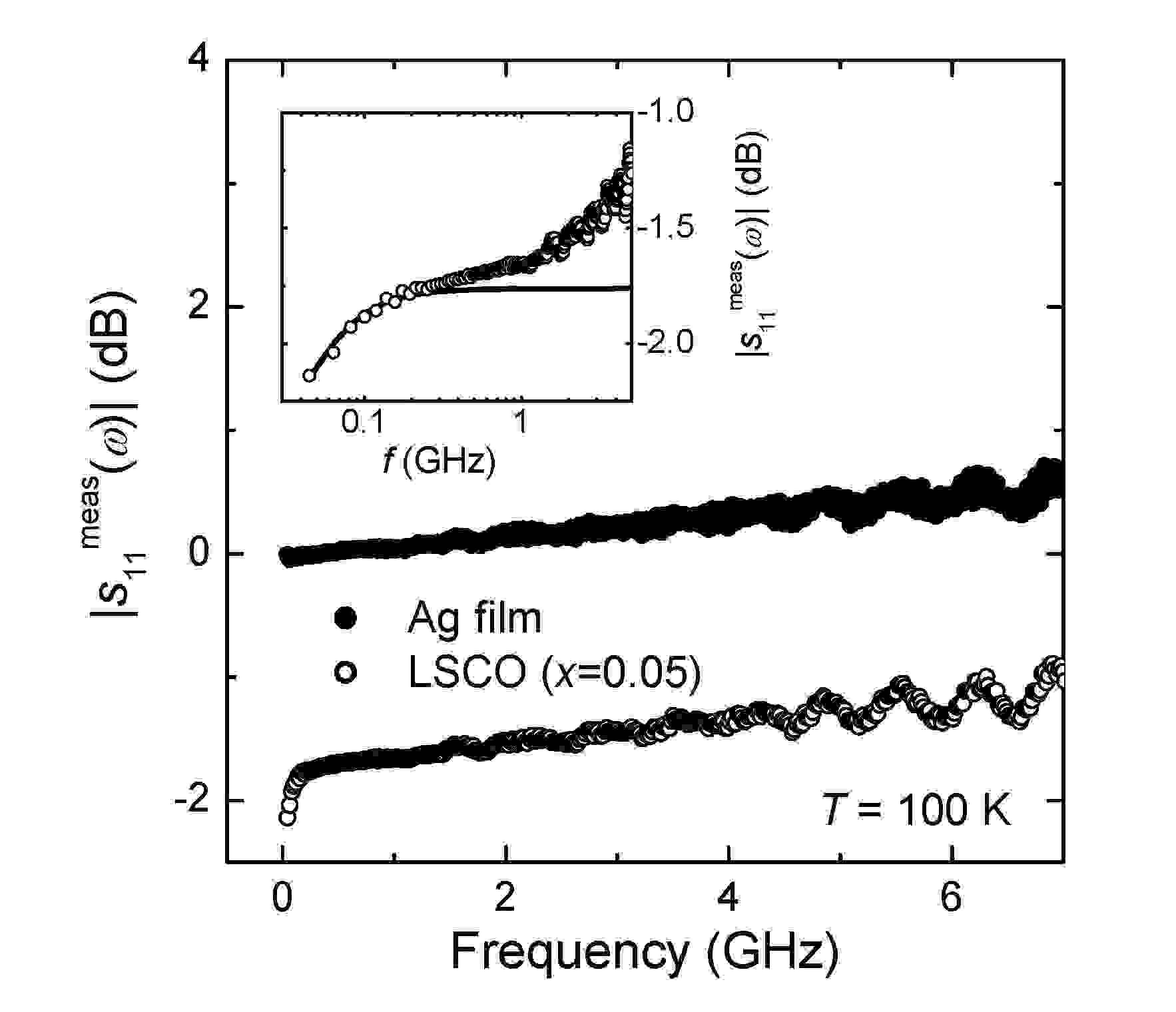}
\caption{
Frequency dependence of $|S_{\rm 11}^{\rm meas}(\omega)|$ at $T$=100~K for a LSCO ($x$=0.05, $t$=616~nm) film with a large contact resistance (open circles) and those for a Ag short film with a negligible contact resistance (solid circles). 
The inset shows the fitting of $S_{\rm 11}^{\rm meas}(\omega)$ for the LSCO ($x$=0.05) sample to an equivalent circuit model drawn in Fig.~\ref{fig:S11_contact} by using $R_c$=5.6~$\Omega$ and $C_c$=0.8~nF. 
}
\label{fig:contact}
\end{figure}

The 3D electromagnetic numerical simulations were performed under the same boundary condition as for our coaxial probe \cite{MW-studio}. 
The similar resonance peaks were observed in $|S_{\rm 11}(\omega)|$, when the relative permittivity of the substrate, $\epsilon$, was assumed to be $\sim$16, as shown in Fig.~\ref{fig:S11_eps_rho}. 
Although the simulation results could not reproduce the same absolute value of $|S_{\rm 11}(\omega)|$ as our experimental results, we captured the following two features showing a good agreement with the experimental results; 
(i) the resonance peaks appearing in $|S_{\rm 11}(\omega)|$ are strongly dependent on the value of $\epsilon$, (ii) they become less prominent by increasing the conductivity of a metallic film. 
In addition, as shown in Fig.~\ref{fig:EMfield}, the plots of the electromagnetic fields at two resonance frequencies clearly suggested that the distribution of the electromagnetic field at $\sim$ 8~GHz (or $\sim$ 23~GHz) agreed with that for the TM$_{\rm 010}$ (or TM$_{\rm 020}$) mode excited in a cylindrical space (diameter is 7mm, height is 0.5 mm) including the dielectric substrate (4$\times$4$\times$0.5 mm$^3$) beneath the metallic film. 

Although the 3D electromagnetic simulation was found to be very powerful for this issue, the numerical results presented here are still preliminary, since the dimension of meshes (typically, $\sim$ 10~$\mu$m) is much larger than the film thickness (typically, $\sim$ 0.1~$\mu$m). 
The main difficulty is that the film thickness is much smaller than the dimension of other components in our coaxial probe. 
More careful simulations optimizing the mesh division are needed to obtain more quantitatively reliable results. 

Based on these results, the influence of the dielectric substrate to the application of Eq.(\ref{eq:sigma}) can be roughly checked by the appearance of this resonance peak. 
As shown in Fig.~\ref{fig:sgm_fluct}, we found that no resonant peak appeared over the frequency range from 0.1 GHz to 12 GHz in the temperature region close to $T_c$, which was consistent with the above-mentioned features. 
We also confirmed that there was no signature of the resonance peak in the raw data of $S_{11}(\omega)$ of the sample at $T=T_0$, where the load standard was adopted in the new calibration method. 
\begin{figure}[t]
\includegraphics[height=5.2cm,width=8cm]{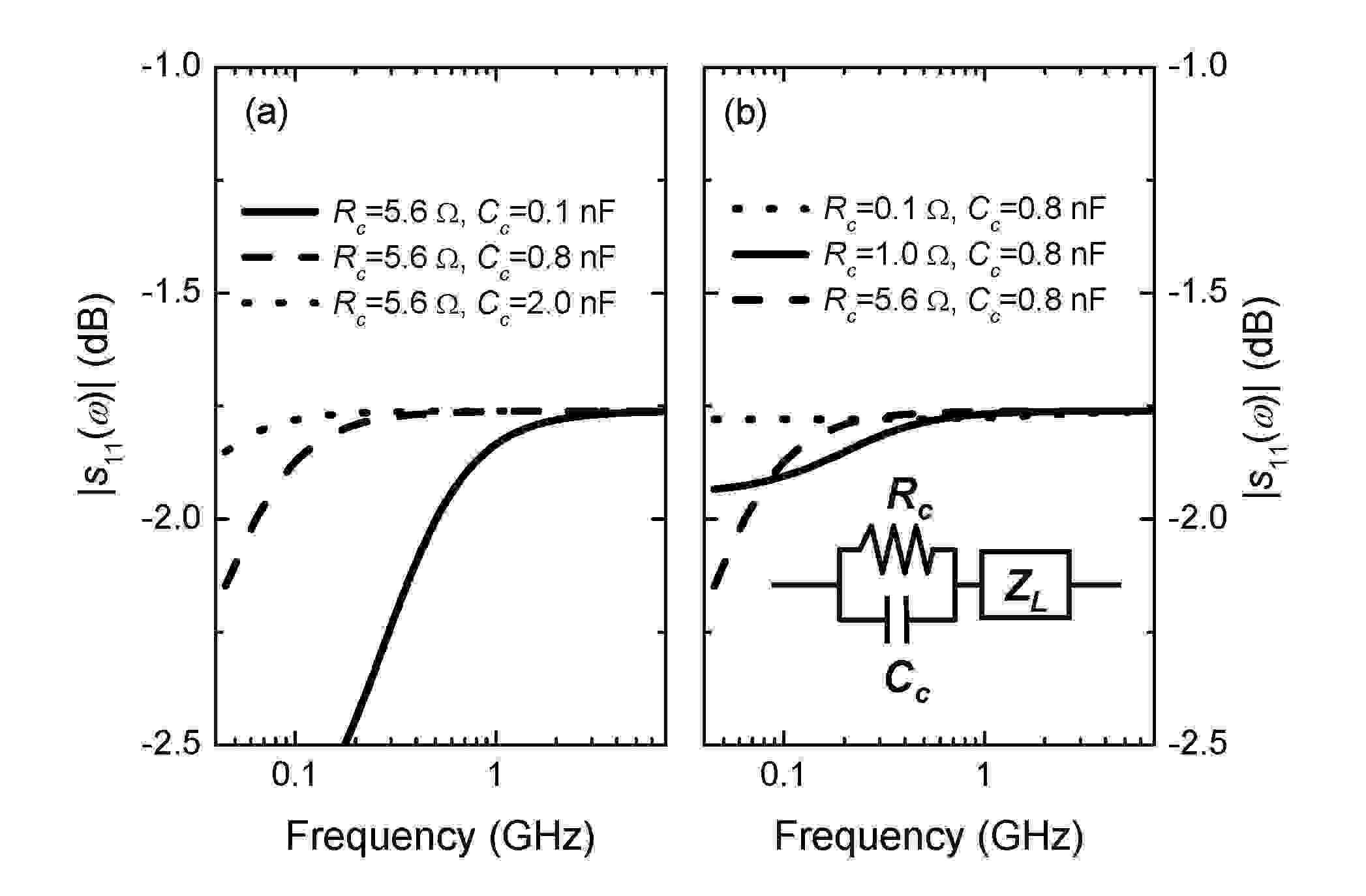}
\caption{
Calculation of $|S_{\rm 11}(\omega)|$ for an equivalent circuit model (drawn in the figure) to consider the influence of the contact electrode, 
by changing the values of (a) $C_c$ and (b) $R_c$, respectively. 
}
\label{fig:S11_contact}
\end{figure}

\section{Influence of the contact electrode}
The microwave broadband measurement described here is one of the two-probed measurement methods, where the contribution of the contact resistance is generally included in the measured data. 
We found that a non-negligible contact resistance gives a sharp dip to the lower frequency part of $|S_{\rm 11}^{\rm meas}(\omega)|$ for a highly conductive thin film sample, as shown in Fig.~\ref{fig:contact}. 
A similar feature was also seen in Fig.~10 of the paper by Scheffler and Dressel~\cite{Scheffler2005}. 
This unusual behavior is understood by a simple equivalent circuit model, where a load impedance ($Z_L$) of the sample is serially connected by a parallel circuit with the contact resistance ($R_c$) and the contact capacitance ($C_c$). 
Thus, $S_{\rm 11}(\omega)$ is given by 
\begin{equation}
S_{\rm 11}(\omega)=\frac{Z_L'(\omega)-Z_c}{Z_L'(\omega)+Z_c}, 
\label{eq:B1}
\end{equation}
where $Z_L'$ is given by 
\begin{equation}
Z_L'=Z_L+\frac{1}{1/R_c+i\omega C_C}. 
\label{eq:B2}
\end{equation}

Figure \ref{fig:S11_contact} shows the calculated results of $S_{\rm 11}(\omega)$ for several values of $R_c$ and $C_c$, suggesting that the contact resistance larger than 1~$\Omega$ and the contact capacitance less than 2~nF give a serious influence to $|S_{\rm 11}(\omega)|$ in the frequency region from 10~MHz to 1~GHz. 
The inset of Fig.~\ref{fig:contact} shows the fitting result of the measured $S_{\rm 11}^{\rm meas}(\omega)$ for a LSCO ($x$=0.05) thin film at $T$=100~K, by using $R_c$=5.6~$\Omega$ and $C_c$=0.8~nF. 
The estimated value of $R_c$ was consistent with the result of dc resistance measurement for the same LSCO film. 
In addition, we found that the estimated value of $C_c$ was comparable to the value for a parallel-plate condenser with the same area as the used corbino electrode, a distance of $\sim$100~nm, and a permittivity of vacuum space, suggesting that the contact capacitance was probably attributed to imperfect adhesion of the corbino electrode. 

The results shown in Fig.~\ref{fig:S11_contact} also suggest that the decrease of $R_c$ and the increase of $C_c$ are very effective to suppress the influence of the contact electrode. 
In the case of the LSCO thin films, they have been made by annealing the samples at 830~$^\circ{\rm C}$ for a few minutes, giving rise to $R_c\sim$ 10~m$\Omega$. 
In order to avoid oxygen deficiency, which may change the superconducting property, the LSCO samples were annealed in O$_2$ atmosphere and were quenched to low temperatures using liquid nitrogen \cite{Ohashi_condmat}. 
On the other hand, any heat treatment cannot be used for the electron-doped LCCO superconductors. 
This is because the T' structure without any apical oxygen, which is crucially important to obtain the superconducting phase in the electron-doped cuprate, is easily damaged by a heat treatment after the crystal growth \cite{Naito2000}. 
Thus, we adopted another method to reduce the contact resistance. 
First, a gold evaporation at room temperature was made in the vacuum chamber used for a film growth, before exposing the film to atmosphere. 
Second, a corbino electrode was fabricated by using the ion-milling technique. 
We confirmed that $R_c$ for LCCO thin films was reduced down to 50~m$\Omega$ by using this method. 

As pointed out in Sec.~IV, the influence of the contact electrode determines the minimum frequency to be measured by the broadband technique. 
Note that the difference of the contact resistance between a sample and a load standard also degrades the lower frequency part of the corrected data. 
It is clear that our calibration method using the normal-state conductivity of the sample as a load standard is free from this problem.



\begin{thebibliography}{70}
\bibitem{coax_cable} The copper and stain-less steel coaxial cables used are manifactured by KEYCOM Corp., parts number are 085E and ULT01, respectively.
\bibitem{small_spring} We used a tiny spring prepared for a contact probe, manufactured by KITA Manufacturing. Corp., parts number U057-516.
\bibitem{memo1} Here, we neglected the contribution of the fringe field excited by the TEM mode inside the coaxial probe and by the TM modes outside the coaxial probe.
\bibitem{memo4} Drude conductivity is expressed by $\sigma_{\rm Drude}(\omega)=\sigma_0/[1+i\omega\tau]$, where $\sigma_0$ and $\tau$ are the dc conductivity and the scattering time, respectively. In the Hagen-Rubens limit ($\omega\tau\gg1$), the real part of $\sigma(\omega)$ is independent of $\omega$ and equal to $\sigma_0$, while the imaginary part of $\sigma(\omega)$ is negligibly small.
\bibitem{Reproducibility}
We note that there was no resonance peak in $S_{\rm 11}^{\rm meas}(\omega)$ for the short and load standards in this case, in contrast to other typical cases. 
This implies a possibility that the reproducibility including the occurrence of the resonance peak governs the maximum frequency where $\sigma(\omega)$ can be successfully obtained. 
As will be discussed in Appendix A, the 3D electromagnetic numerical simulations are considered to be very useful for resolving this issue.
\bibitem{memo5} Indeed, the highest magnetic field in this work was limited by the applicability of our calibration scheme, which was found to be dependent on materials investigated. In the case of LSCO, we confirmed that our scheme was useful up to $B$=1~T.
\bibitem{MW-studio} CST, MICROWAVE STUDIO, ver.5. 
The numerical simulation was performed in the transient solver mode. 
We calculated $S_{\rm 11}(\omega)$ under the same boundary condition as for our coaxial probe. The typical mesh size was 1.5$\times$10$^5$.
\end{thebibliography}
\end{document}